\documentclass[epsfig,graphics,floatfix,nofootinbib,notitlepage,twocolumn]{revtex4-1}
\usepackage[table]{xcolor}
\usepackage{amsmath,amsfonts,amssymb,graphics,graphicx,epsfig,color,times,bbm}
\usepackage{amsthm}
\usepackage{mathtools}
\usepackage{psfrag}
\usepackage{braket}
\graphicspath{{figures/}}
\usepackage[hidelinks]{hyperref}
\hypersetup{colorlinks=false}
\usepackage{wasysym}
\usepackage{bbm}

\usepackage{hyperref}
\usepackage{lipsum}
\usepackage{booktabs}

\usepackage{changes}
\definechangesauthor[name={Sam Roberts}, color=purple]{sr}
\definechangesauthor[name={Karthik Seetharam}, color=blue]{ks}
\usepackage[normalem]{ulem}

\begin{document}


\title{Fault-tolerant Post-Selection for Low Overhead Magic State Preparation}
\author{H\'ector Bomb\'in}
\author{Mihir Pant}
\author{Sam Roberts}
\author{Karthik I. Seetharam\footnotemark[0]} \email{Lead author: karthik@psiquantum.com}

\affiliation{PsiQuantum, Palo Alto}

\date{\today}

\begin{abstract}

We introduce a framework for fault-tolerant post-selection (FTPS) of fault-tolerant codes and channels---such as those based on surface-codes---using soft-information metrics based on visible syndrome and erasure information. We introduce several metrics for ranking configurations of syndromes and erasures. 
In particular, we introduce the \emph{logical gap} (and variants thereof) as a powerful soft-information metric for predicting logical error rates of fault-tolerant channels based on topological error-correcting codes. The logical gap is roughly the unsigned weight difference between inequivalent logical corrections and is adaptable to any tailored noise model or decoder.
We deploy this framework to prepare high-quality surface code magic states with low overhead under a model of independent and identically distributed (\emph{i.i.d.}) Pauli and erasure errors. Post-selection strategies based on the logical gap can suppress the encoding error rate of a magic state preparation channel to the level of the physical error rate with low overhead. For example, when operating at $60\%$ the bulk threshold of the corresponding surface code, an overall reduction of the encoding error rate by a factor of $15$ is achievable with a relative overhead factor of ${< 2}$ (approximately $23$ times less than that of simple syndrome-counting rules). We analyze a schematic buffer architecture for implementing post-selection rules on magic state factories in the context of magic state distillation. The FTPS framework can be utilized for mitigating errors in more general fault-tolerant logical channels. 

\end{abstract}

\maketitle

\section{Introduction} \label{sec:Intro}

Post-selection is an essential ingredient in many universal schemes of fault-tolerant quantum computation. For fault-tolerant architectures based on 2D topological stabilizer codes such as surface codes~\cite{kitaev1997quantum, bravyi1998quantum, dennis2002topological, kitaev2003fault, kitaev2006anyons, bombin2009quantum, bombin2010topological, raussendorf2007fault,raussendorf2007topological, horsman2012surface, hastings2014reduced, terhal2015quantum, brown2017poking, litinski2019game, bombin2021logical} (and related approaches~\cite{bombin2006topological, landahl2011fault, barkeshli2013classification, barkeshli2013twist, yoder2017surface, bombin2018transversal, bombin20182d, lavasani2018low, lavasani2019universal, webster2020fault, roberts20203, zhu2021topological,chamberland2021universal,landahl2021logical}), it is ubiquitous; in order to perform logical non-Clifford gates, magic state distillation and injection are used~\cite{bravyi2005universal, bravyi2012magic, fowler2013surface, haah2017magic, campbell2017unified, haah2018codes, gidney2019efficient, litinski2019magic, holmes2019resource}, heavily utilizing post-selection in the process. In particular, to prepare magic states that are required for injection, many noisy magic states are fed into a magic state distillation protocol, producing fewer magic states of significantly higher quality as a result of post-selection---the high-quality magic states are only output if certain error-detecting measurements in the protocol do not flag the presence of an error.

For a given protocol, the total overhead of magic state distillation is strongly dependent on the quality of the initial noisy magic states. For example, to first order, the well known 15-to-1 distillation protocol takes initial magic states with error rate $p$, and produces fewer magic states with error $35p^3$~\cite{bravyi2005universal, bravyi2012magic}. If $p<10^{-3}$, to reach a target logical error rate per logical operation of $10^{-14}$ as is needed in, for example, quantum chemistry applications~\cite{kivlichan2020improved, von2020quantum, kim2021faulttolerant, su2021faulttolerant}, one typically needs to iterate this 15-to-1 procedure twice (i.e., two rounds of distillation). An additional round is needed if the physical error rate is $p=10^{-2}$. Hence, the initial magic state quality can severely affect resource overhead, with a large penalty incurred every time an additional round is added. This is under an idealized model with perfect Clifford gates utilized in the distillation protocol. 

As both the magic states and gates in the distillation protocol are imperfect, they will be encoded in a quantum error-correcting code---such as the surface code~\cite{kitaev1997quantum, bravyi1998quantum, dennis2002topological}. These additional sources of noise reduce the performance of distillation and need to be accounted for to determine the overhead and output magic state(s) logical error rate(s).
For instance, encoding a magic state in a quantum error-correcting code, such as the surface code, introduces additional \textit{encoding error} that can be mitigated (using post-selection strategies) in accordance with the desired overall error rate for a noisy encoded magic state. Hence, when estimating (or optimizing) the overhead for distillation, one should also include the cost for preparing the initial magic states at a desired quality.
Initial work by Li~\cite{Li2015MSP} demonstrates a reduction in the encoding error rate under circuit level noise using simple post-selection scheme based on syndromes. In particular, the magic state preparation is only accepted if no syndromes are observed after a few rounds of stabilizer measurements. This protocol produces low error rate magic states (particularly when the noise model is dominated by two-qubit errors), although a general analysis of encoding error vs. overhead is absent. Singh et al.~\cite{singhpuri2021MSPbiased} modify the no-syndrome post-selection protocol of Li by encoding the initial magic state in an error-detecting code, yielding a reduction of the overall preparation block error rate under a biased noise model.

In this paper, we introduce a general framework for \emph{fault-tolerant post-selection} (FTPS) of surface code channels (also known as logical blocks in Ref.~\cite{bombin2021logical}) along with several efficient rules for post-selection based on soft information obtained from the visible syndrome and erasure. We apply these rules to the problem of preparing magic states encoded in the surface code. In particular, we find that for post-selection rules based on the \emph{logical gap} and its derivatives, we can improve the quality of the initial encoded magic states by suppressing the encoding error rate by orders of magnitude with modest additional overhead, under an independent and identically distributed (\textit{i.i.d.}) model of Pauli and erasure errors, and over a wide range of error rates. For example, we see that when the physical error rate is approximately $60\%$ of the bulk threshold, we can suppress the encoding error rate to that of the physical error rate using a post-selection overhead of less than $2$. This constitutes an overall reduction of the encoding error rate by a factor of $\sim15$, leading to significant resource savings in the overall magic state distillation protocol.

\section{Magic State Preparation} \label{sec:MagicStatePrep}

\begin{figure*}
	\centering
    \includegraphics[width=0.9\linewidth]{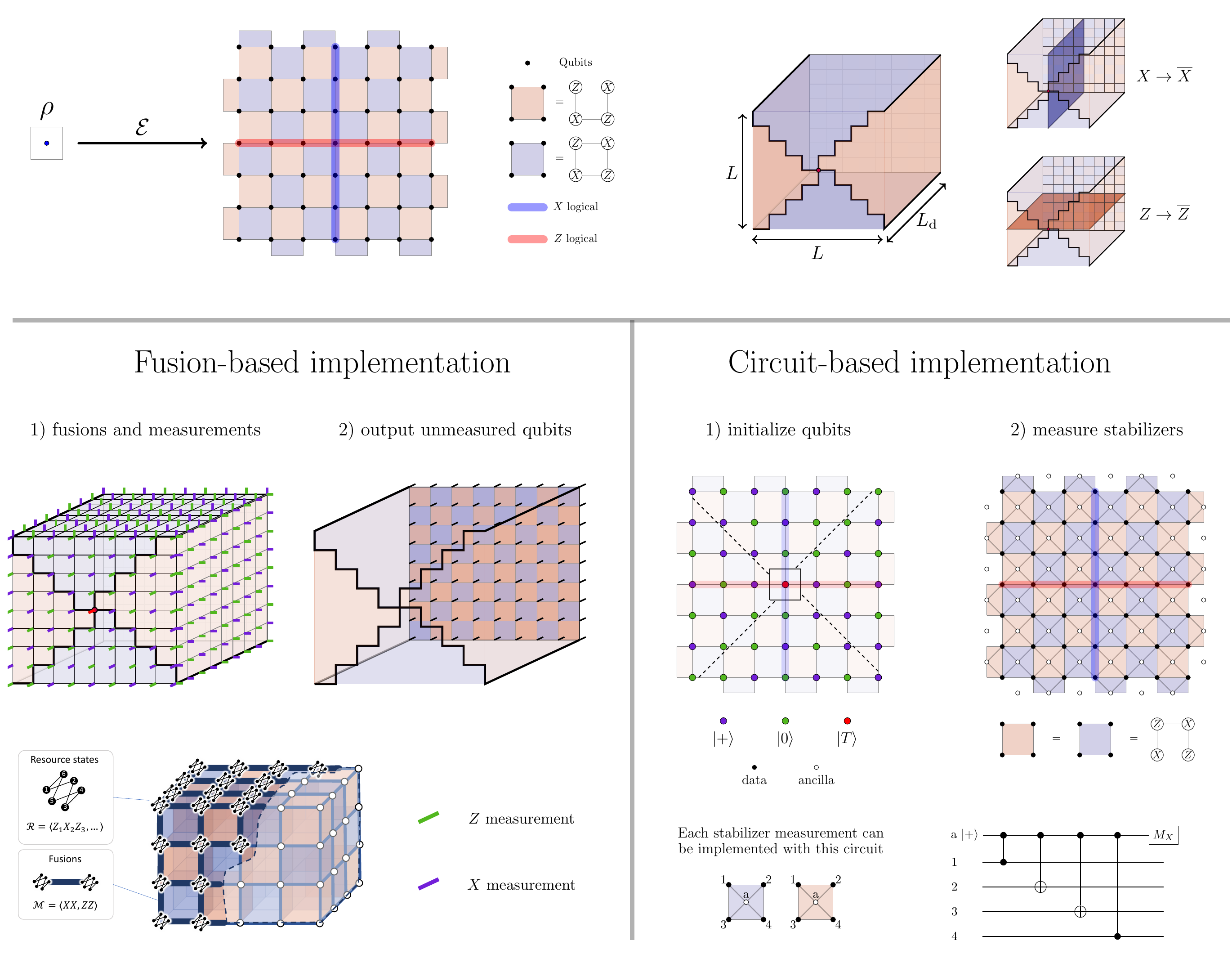} 
	\caption{ (top left) Noisy magic state preparation can be regarded as an channel taking a distance-$1$ code state (i.e., an unencoded state), to a distance-$L$ code state. (top right) The space-time diagram for the location of boundaries, where blue and red denote the two distinct boundaries. The noisy initial magic state supported on the central qubit on the front is encoded in a surface code on the rear. We remark that the the surface code on the boundary has a local basis that depends on the orientation of the boundary (e.g., it may be the $ZXXZ$ or $XZZX$ version of the surface code). The membrane showing how the $X$ ($Z$) operator of the initial magic state is mapped to logical $\overline{X}$ ($\overline{Z}$) operator on the surface code is shown in blue (red). 
	(bottom left) The measurement pattern for implementation in FBQC using the $6$-ring fusion network~\cite{bombin2021logical,bombin2021interleaving}. Individual qubits belonging to resource states on the boundary are measured in the $X$ ($Z$) basis as accordingly depicted by purple (green) edges. To prepare an encoded $\ket{T} = \frac{1}{\sqrt{2}}(\ket{0} \pm e^{\frac{i \pi}{4}} \ket{1})$ on the output, the qubit belonging to the central resource state is measured in the $\frac{1}{\sqrt{2}}(X + Y)$ basis; the $\pm$ sign is determined by the measurement outcomes. (bottom right) The initial configuration of qubits for a circuit-based implementation with a planar array of qubits. Measuring the surface code stabilizers implements results in the desired encoding.
	}
	\label{fig:MagicStatePreparation}
\end{figure*}

To perform a distillation protocol with fault-tolerant gates, we require the input (noisy) magic states to be encoded. Here, we are specifically interested fault-tolerant computations based on the surface-code~\cite{kitaev1997quantum, bravyi1998quantum, dennis2002topological}. The surface code is a stabilizer code~\cite{gottesman1997stabilizer,gottesman2010introduction}, meaning it is defined by an abelian subgroup $\mathcal{S}$ of the Pauli group $\mathcal{P}_n$ (on $n$-qubits), not containing $-I$. Here, we consider the Wen version~\cite{wen2003quantum} (or $ZXXZ$-version~\cite{kay2011capabilities}). It is defined by placing a qubit on the vertices of a square lattice, with one stabilizer generator per plaquette, formed as a product of Pauli $ZXXZ$ on the four qubits in its support. By introducing boundaries of the code, as depicted in Fig.~\ref{fig:MagicStatePreparation}, the code defines one logical qubit, with logical operators $\overline{X}$ and $\overline{Z}$ defined as strings of Pauli operators spanning opposite boundaries (also depicted in Fig.~\ref{fig:MagicStatePreparation}). 

The preparation of these noisy magic states can be phrased as an encoding problem. Namely, letting $Q$ be the state space of the noisy \emph{initial magic state} qubit, and $X$ and $Z$ the single qubit Pauli operators acting on it. We define a protocol to implement the following encoding isometry:
\begin{equation}
    \mathcal{E}: Q \rightarrow Q^{\otimes d^2}, \text{ such that }  X \mapsto  \overline{X}, ~ Z \mapsto  \overline{Z}.
\end{equation}
where $\overline{X}$ and $\overline{Z}$ are the logical operators of the surface code. We outline two approaches to achieve this encoding isometry---one based on fusion-based quantum computation (FBQC) with the $6$-ring fusion network~\cite{bartolucci2021fusion}, and one based on circuit-based quantum computation (CBQC) with a planar array of qubits. These approaches build upon the proposal of \cite{lodyga2015simple} and are also discussed in \cite{brown2020universal, bombin2021logical}.

\subsection{Preparation protocol} \label{subsec:PrepBlock}

We follow the construction in \cite{bombin2021logical}. The protocol contains two parameters, $L$ and $L_{\mathrm{d}}$. We refer to $L$ as the ``distance" of the scheme -- it determines the code distance of the surface code state we are preparing. We refer to $L_{\mathrm{d}}$ as the ``depth" of the scheme -- it can be thought of as simulated time, i.e., the number of rounds of stabilizer measurements in CBQC, or the number of layers of resource states in FBQC, and determines the number of stabilizer checks in the protocol from which we can gather soft-information for post-selection. One may choose a minimal depth of $L_{\mathrm{d}}=2$, as is done in \cite{Li2015MSP, singhpuri2021MSPbiased}, allowing the encoded magic states to be prepared. However, we consider longer depths (which requires more temporal overhead in both FBQC and CBQC for the construction of a single block\footnote{Here we study the post-selection overheads (for different rules) relative to the construction of a single block and do not compare overheads of post-selection strategies on short-depth and larger-depth blocks.}) to allow for more information to be collected in order to better predict logical errors on the output state.

\textbf{Circuit-based protocol.} In CBQC, the preparation protocol is described in Fig.~\ref{fig:MagicStatePreparation} (right). We begin with a 2D $L\times L$ array of qubits. One of these qubits is prepared as (a noisy version of) the initial magic state $\ket{T}$, while the remaining qubits are prepared in an eigenstate of Pauli-$X$ or $Z$ according to the figure. We then perform $L_{\mathrm{d}}$ repeated rounds of surface-code stabilizer measurements. The state after these measurements is an encoded version of the initial magic state qubit.

\textbf{Fusion-based protocol.} In FBQC \cite{bartolucci2021fusion, bombin2021logical}, the bulk of the preparation block consists of 6-ring resource states that are fused along a cubic lattice of size $L\times L\times L_\mathrm{d}$, with each pair of qubits from adjacent resource states in each of the three orthogonal directions undergoing a two-way fusion, i.e., a Bell measurement (e.g., $XX$ and $ZZ$ measurements) as in Fig.~\ref{fig:LogError}. Boundaries are formed by single qubit measurements in an alternating $X$ and $Z$ pattern, with the distinction between primal (blue) and dual (red) boundaries given by a translation of the alternating pattern by one site (or alternatively, flipping the $X$ and $Z$ measurements) as in Fig.~\ref{fig:MagicStatePreparation} (bottom left). There is redundancy among the measurement outcomes; certain measurements may be multiplied together to form a \textit{check} operator, whose outcome can be used to detect errors. More precisely, check operators are elements of both the (joint) stabilizer group of the resource states as well as the measurement group (which includes fusions and boundary measurements)~\cite{bartolucci2021fusion}. One may multiply the measurement outcomes comprising a check to construct the \textit{syndrome}---in the absence of error, these syndrome measurements should have even parity, and as such, an odd parity signals the presence one or more errors. 

To complete the protocol, a single qubit in the resource state at the preparation point on the input port is measured in the magic state basis $\frac{1}{\sqrt{2}}(X+Y)$. This yields an initial magic state $T\ket{\pm}$ qubit that is entangled with the rest of the block via the bulk two-way fusions (where $\ket{\pm}$ is the $\pm$ eigenstate of $X$, and is determined by the measurement outcome of the $\frac{1}{\sqrt{2}}(X+Y)$ measurement). The output of this channel is an encoded (noisy) $\ket{T}$ state on surface code supported on the remaining unmeasured qubits, up to a Pauli operator depending on fusion and measurement outcomes.

\textbf{Space-time diagram.} An abstract space-time diagram of this channel used to achieve $\mathcal{E}$ is depicted in Fig.~\ref{fig:MagicStatePreparation} (top right). In particular, time can be thought of as running into the page with the noisy initial magic state situated in the center of the the input port (initial time slice), which we call the \emph{preparation point}, and with the encoded magic state supported on the output port (final time slice). The operator $X$ ($Z$) on the input port is mapped to $\overline{X}$ ($\overline{Z}$) on the output port via the logical membrane. Here, following \cite{bombin2021logical}, a logical membrane is the world-sheet of a logical operator. It specifies how the input and output logical operators are correlated. 

\subsection{Distances and logical errors} \label{subsec:DistancesErrors}

\begin{figure*}[t]
	\centering
    \includegraphics[width=0.22\linewidth]{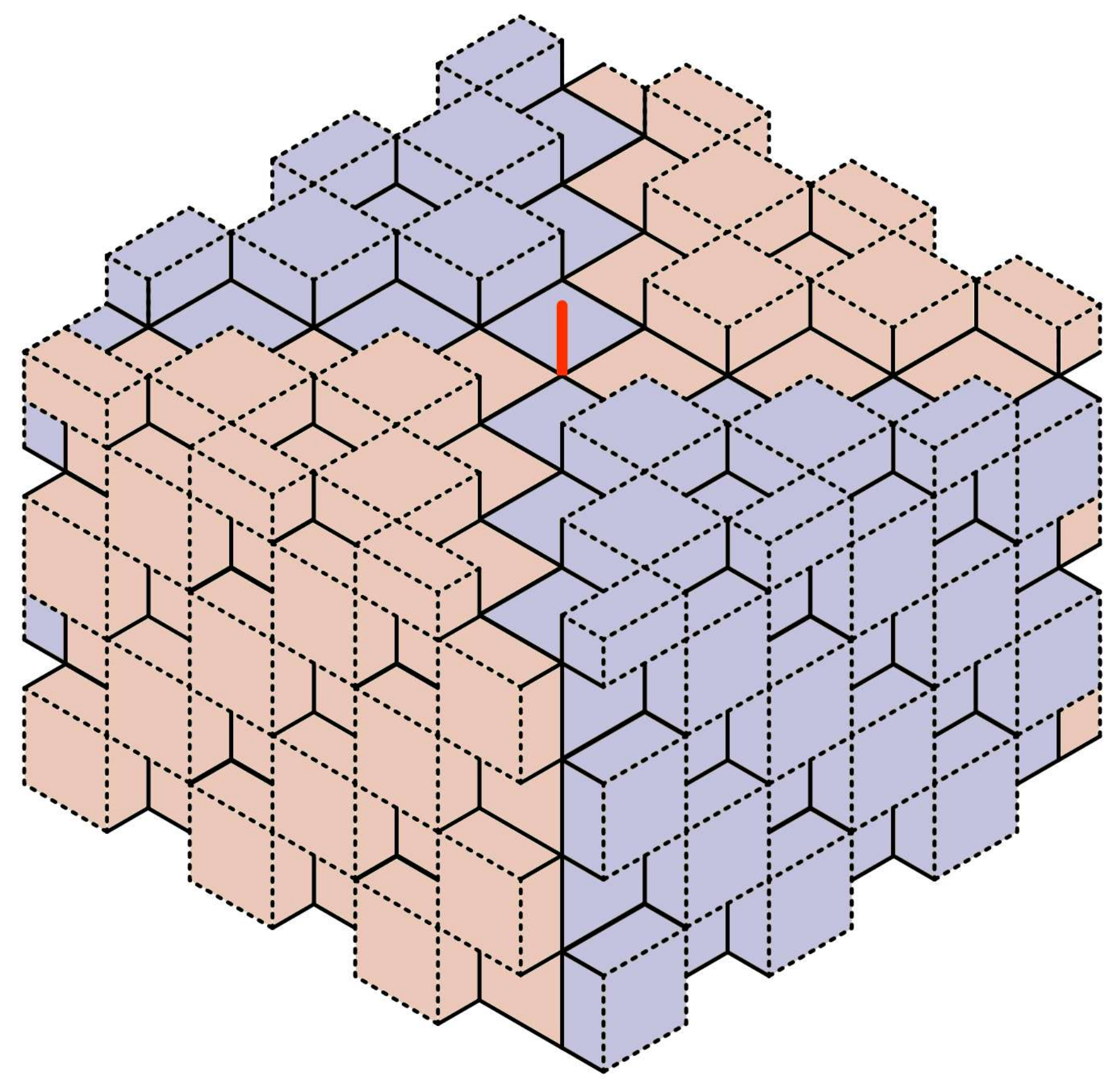} \qquad 
    \includegraphics[width=0.18\linewidth]{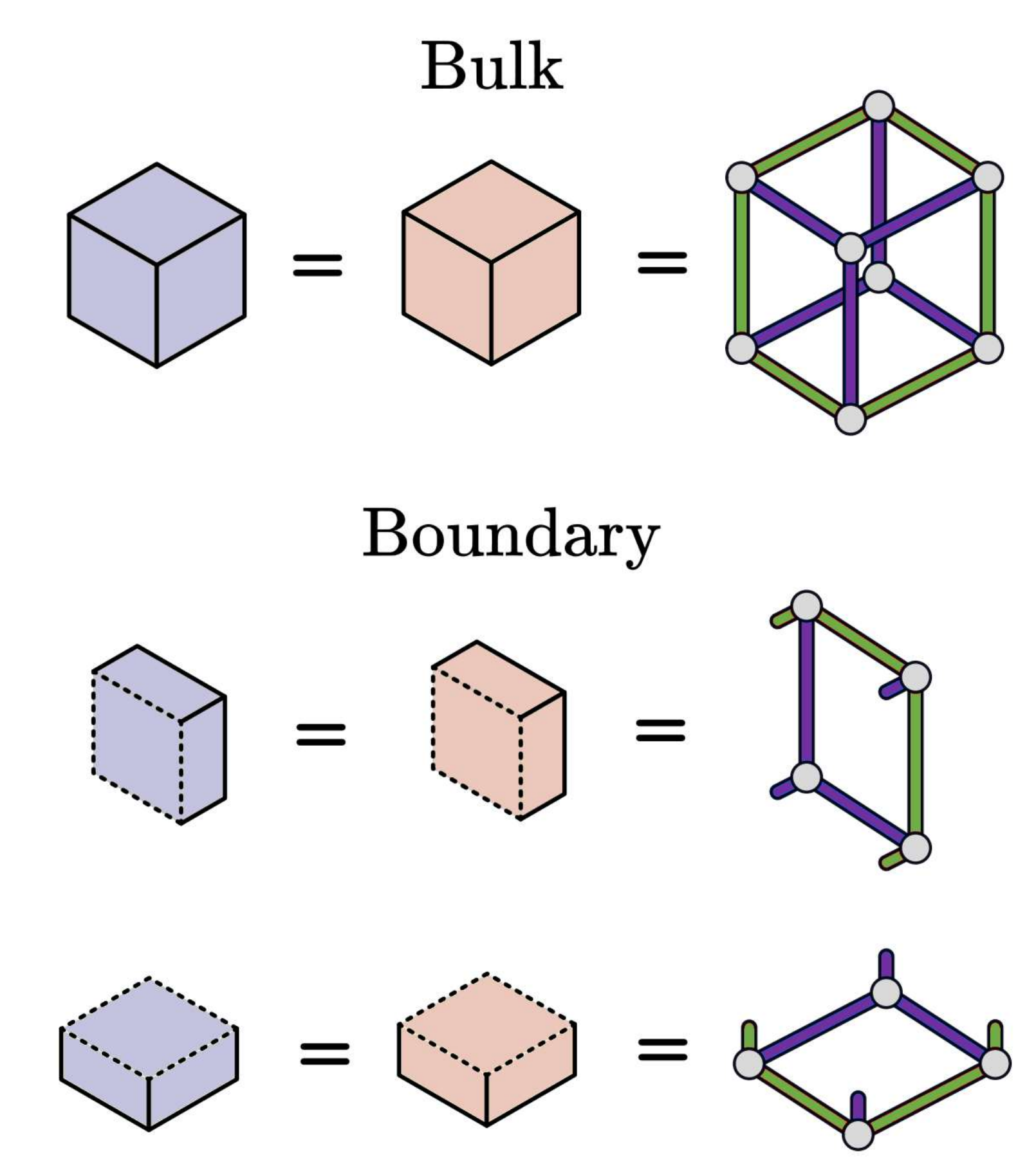} \qquad \qquad
    \includegraphics[width=0.28\linewidth]{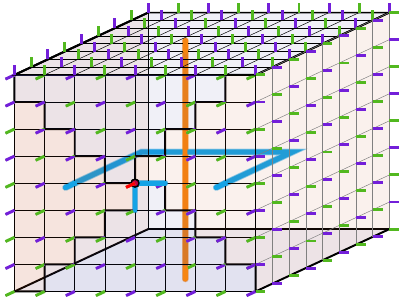}
	\caption{
	(left)  The check operator structure of the preparation protocol in the 6-ring fusion network. There is a check operator for each cube of the fusion network (suitably truncated for boundaries). Here, the block has a different orientation (and as such, the measurement pattern is slightly different), such that ``time'' flows from top to bottom. (right) Examples of logical errors for the preparation protocol. Chains of flipped fusion outcomes between distinct boundaries lead to logical errors. In particular, there are weight-2 logical errors supported near the central magic-state. Red (blue) error chains flip red (blue) check operators and logical membranes.
	}
	\label{fig:LogError}
\end{figure*}

The fault-distance of the protocol we present is constant, as there is a space-time volume around the preparation point where low-weight errors can give rise to logical errors. In particular, for the FBQC protocol, barring the initial magic state measurement itself, the fault distance is $2$; minimally, two fusion outcomes neighbouring the initial magic state measurement can be flipped in an undetectable way, yielding a logical error. Such a minimal error is shown in Fig.~\ref{fig:LogError} in addition to other representative non-trivial logical errors. In CBQC (or measurement-based quantum computation), the corresponding protocol has a fault distance of 3, meaning 3 single qubit Pauli errors can introduce a logical error (see, for example, Fig 13. of Ref.~\cite{brown2020universal}). In principle, a depth of 2 is sufficient to produce an encoded magic state. In practice, however, choosing a larger depth provides more syndrome information to more reliably detect and correct such errors.

\subsection{Error Model} \label{subsec:ErrorModel}

It is helpful to separate the overall preparation error into the error on the initial magic state and the error on the remainder of the channel. Assuming both of these errors occur independently, the
overall error of the magic state preparation block becomes
\begin{equation} \label{eq:preperror}
 p_{\mathrm{prep}}=p_{\mathrm{init}}(1-p_{\mathrm{enc}})+(1-p_{\mathrm{init}})p_{\mathrm{enc}}   
\end{equation}
where $p_{\mathrm{init}}$ is the error rate on the initial magic state (measurement)
and $p_{\mathrm{enc}}$ is the error rate on the remainder of lattice that
encodes the magic state in a surface code on the output port; we call the latter the encoding error rate (EER). To leading
order in the low error regime, $p_{\mathrm{prep}}\approx p_{\mathrm{init}}+p_{\mathrm{enc}}$.
We consider the situation where $p_{\mathrm{init}}$ is fixed, typically
by physical hardware and design choices in the architecture,
and focus on reducing $p_{\mathrm{enc}}$ via post-selection, which as we will see later, is by far the dominant source of error contributing to $p_\mathrm{prep}$ in the absence of FTPS.\footnote{One may accomplish additional reduction of $p_\mathrm{init}$ with minimal overhead using hardware optimizations specifically tailored to the preparation point.} Therefore,
in the following, we ignore $p_{\mathrm{init}}$ and consider only
$p_{\mathrm{enc}}$ as the logical error rate on the encoding lattice
arising from \textit{i.i.d.} erasure and Pauli errors on every edge
of the lattice. In FBQC, this error model corresponds to each measurement in a fusion
suffering an erasure with probability $p_{\mathrm{erasure}}$ or suffering
a pauli error with probability $p_{\mathrm{error}}$ conditioned on
not being erased. In CBQC, this error model corresponds to an erasure
error on each qubit (or measurement), e.g., arising from a leakage error, or a Pauli error on each qubit (or measurement) conditioned on not being erased. In the main text, we will discuss only Pauli errors, relegating the discussion the mixed erasure and Pauli errors to Appendix \ref{app:erasures}. In a similar vein, by Eq.~\ref{eq:preperror}, decreasing $p_\mathrm{init}$ by improving the quality of the initial magic states, while not mitigating $p_\mathrm{enc}$ will also lead to diminishing returns when $p_\mathrm{init}\ll p_\mathrm{enc}$. Hence, it is desirable to mitigate both sources of error. Here we address how to systematically suppress the EER $p_\mathrm{enc}$, given a fixed $p_\mathrm{init}$.

Given the constant fault-distance of $2$ when ignoring the initial magic state measurement in the FBQC preparation block, even with post-selection, we cannot hope to reduce the encoding error rate $p_\mathrm{enc}$ of the channel to below $\mathcal{O}(p_\mathrm{error}^2)$. In practice, this is not a bottleneck, as $p_\mathrm{prep}$ will always limited by the quality of the initial magic state $p_\mathrm{init}$, which is be proportional to the single qubit error rate $p_\mathrm{error}$. Our goal is to reduce the encoding error rate $p_\mathrm{enc}$ by as much as possible.

\section{Fault-tolerant Post-selection} \label{sec:FTPS}

In this section, we introduce the general framework of fault-tolerant
post-selection and define a set of post-selection rules for encoding magic
states in surface codes. 
For a given logical block $B$ (in any model of computation, CBQC, FBQC, or measurement-based quantum computation), we define a \textit{block configuration} $E$ as a set of Pauli errors $\epsilon$ and erasure errors $\varepsilon$ on $B$. Given the check operators of the logical block (e.g., those of the 6-ring in Fig.~\ref{fig:LogError}), we can deduce the syndrome $\sigma$. The combined information of the syndrome and the erasure information is collectively called the \textit{visible information} $v_{E}=(\sigma,\varepsilon)$. We let the space of all visible information for a given block be denoted $V_{B}$. Here, the logical block is the magic state preparation block, which has the parameters $(L,L_{\mathrm{d}})$.

A post-selection \emph{rule }$R$ observes the visible information
$v_{E}$ and decides whether to accept or reject the block with configuration $E$ using a \emph{soft-information
function $Q$} followed by a \emph{policy $P$}:
\begin{enumerate}
\item A soft-information function $Q:V_{B}\rightarrow\mathbb{R}^{q}$
maps the visible information $v_{E}\in V_{B}$ to a vector of
soft-information data $q_{E}\in\mathbb{R}^{q}$. This step distills
useful and actionable information about $B$ based on $E$. 
\item A policy $P:\mathbb{R}^{q}\rightarrow\{0,1\}$ digests the soft-information
$q_{E}$ and produces a decision on whether to accept $(1)$ or reject
$(0)$ the block $B$ based on the configuration $E$. In general, the policy can be any function of choice. Often, however, this is achieved by a \emph{scoring
function $S:\mathbb{R}^{q}\rightarrow\mathbb{R}^+$} that maps $q_{E}$
to a numerical score for the block, from which the binary decision
is achieved by accepting blocks below a certain cutoff score, i.e., $P=\Theta(s^{*}-S(q_{E}))$,
where $\Theta$ is the Heaviside function, and $s^{*}$ is a cutoff
score such that all configurations with $S(q_{E})\leq s^{*}$ are
kept. Examples of both cases will be shown in Sec.~\ref{subsec:Rules}.
\end{enumerate}

For brevity, we will often write $S(Q)$ to refer to the function $S\circ Q$ that returns a score for some visible information. If, on average, $\kappa$ fraction of blocks are kept, then
the post-selection rule has an average resource overhead of $O\coloneqq\frac{1}{\kappa}$
times the overhead of creating a single block (see Sec.~\ref{subsec:ArchDesign} for more details). The goal is to construct a rule such that the logical error rate (determined via decoding) on the subset of the $\kappa$
accepted blocks is significantly less, on average, than that
on all blocks. This occurs when $R$ strongly correlates the
policy output (often achieved through the score $S(Q)$) with the likelihood of logical error, thereby
facilitating easy selection of less-error-prone blocks. Furthermore,
a high-performing rule in practice would also have low overhead. A rule that is high-performing in
terms of error suppression but requires large, potentially exponential, overhead
is likely impractical beyond small block sizes.

\subsection{Rules} \label{subsec:Rules}

We now define several rules which we name: annular syndrome, logical gap, nested logical gap, and radial logical
gap. We discuss an additional rule we name the surviving distance in App.~\ref{app:survivingdist}. These rules can be applied to \textit{any} block and configuration but here
we tailor the rules towards the problem of magic state preparation,
a natural setting for applying post-selection techniques. We will further focus on FBQC with the 6-ring network for concreteness, but the techniques readily generalize to other models and schemes.

In these definitions, and the simulations that follow, we make use of the notion of a \textit{syndrome graph}. The syndrome graph is defined by placing a vertex for each check operator (bulk cubes and boundary checks) of the fusion network. We connect two vertices with an edge whenever the corresponding check operators utilize a common measurement outcome. For the 6-ring fusion network, there are two distinct syndrome graphs termed the primal/dual syndrome graphs, analogous to the planar surface code, with the bicolorability in Fig.~\ref{fig:MagicStatePreparation} indicating checks belonging to the two independent syndrome graphs, i.e., neighbouring vertices associated to blue (red) checks are connected with an edge, forming the primal (dual) syndrome graph. Furthermore, the magic state preparation block has only $2$ logical membranes, one supported on the edges of each syndrome graph. 

For more general logical blocks encoding channels from $m$ to $n$ qubits, there are $m+n$ independent logical membranes that generate all possible logical correlations from input to output (see Ref.~\cite{bombin2021logical} for more details). We denote the set of independent logical membranes by $C$, which index $2^{m+n}$ logical sectors.

\subsubsection{Annular Syndrome} \label{subsubsec:AnnularSyndrome}

The annular syndrome rule $R_{\mathrm{S}}=(Q_{\mathrm{S}},P_{\mathrm{S}})$
relies solely on syndrome information. It computes the weighted
sum of the $-1$ (``lit up'') syndromes. We choose the weights according to a power-law decay from the preparation point. The intuition
is that syndromes near the preparation point are more significant
in predicting a logical error than those further in the bulk. As such, the soft-information function maps to a vector of length $2$ with the components
\begin{align}
    Q_{\mathrm{S},i}(v_{E};\alpha)\coloneqq\sum_{r=1}^{L_{\mathrm{d}}}\frac{\sigma_{\mathrm{i}}(r)}{\bar{\sigma}_{i}(r)\min(r,\lceil3L/4\rceil)^{\alpha}},i=\mathrm{primal,\;dual}
\end{align}
where $\sigma_{i}(r)$ is defined as the total number of $-1$ syndrome
outcomes at a distance of $r$ from the initial
magic state, termed an ``annulus'' with radius $r$, $\bar{\sigma}_{i}(r)$ is the total number of syndrome measurements (independent of outcome) in the same annulus
of radius $r$, and $\alpha$ is a tunable parameter. One can choose any metric to define the radius; here we use the $L_\infty$ metric (also known as the supremum metric) on the fusion network (depicted in the bottom left of Fig.~\ref{fig:MagicStatePreparation} where resource states reside on vertices of the cubic lattice)\footnote{For the preparation block, is equivalent to the graph distance on each of the syndrome graphs.}. We apply a radial cutoff of $\lceil\frac{3L}{4}\rceil$ to ensure that for large depth blocks where $L_\mathrm{d}>L$, there is no tail region at large radius where syndromes are counted with almost no weight, i.e., there must be some minimum penalty for having syndromes. Note that for more general topological codes one may not have a split primal and dual syndrome graph structure, and one can simply sum over all syndromes in a radius around the preparation point.

The policy is implemented by thresholding a score

\begin{align}
S_{\mathrm{S}}(Q_{\mathrm{S}})&\coloneqq\sum_{i=\mathrm{primal,\;dual}}a_{i}Q_{\mathrm{S},i} \nonumber\\
P_{\mathrm{S}}(Q_{\mathrm{S}};s_{\mathrm{S}}^{*}) &\coloneqq\Theta(s_{\mathrm{S}}^{*}-S_{\mathrm{S}}(Q_{\mathrm{S}}))
\end{align}
where $a_{i}$ are tunable linear weights to construct a combined score
from the primal and dual graph annular syndromes.

\subsubsection{Logical Gap}  \label{subsubsec:LogicalGap}

The logical gap rule $R_{\mathrm{G}}=(Q_{\mathrm{G}},P_{\mathrm{G}})$
is inspired by the statistical mechanical mapping of error correction
in \cite{dennis2002topological} whereby an error correction threshold is equivalent to the phase transition in a related statistical mechanical model. Above the threshold, logical errors are not suppressed due to a loss
of distinguishability between distinct logical sectors. In other
words, above the threshold, the decoder can no longer reliably differentiate which logical sector of the code space to recover to (as the code distance increases).
In this spirit, one can define the logical gap as the difference between
the correction weights that return the system to different logical
sectors. 

For example, in the simple case of a single logical $\bar{Z}$
operator in a surface code memory block (e.g., only the primal syndrome
graph), with a configuration $E$ and possible corrections $\bar{l}_{\mathrm{correct}}, \bar{l}_{\mathrm{wrong}}$ such that composing the correction and error yields a logical operator on the code space---namely $\bar{I}$ and $\bar{Z}$, respectively. The \textit{signed logical gap} is defined as
\begin{align}
    \Delta_{\bar{Z}}(E)\coloneqq w_{\bar{Z}}(\bar{l}_{\mathrm{wrong}})-w_{\bar{Z}}(\bar{l}_{\mathrm{correct}})
\end{align}
where $w_{\bar{Z}}(\bar{l})$ denotes the log-likelihood weight of
the correction $\bar{l}$ for the $\bar{Z}$ sector given by a choice
of decoder, defined as follows: an edge $e$ has weight $w_{e}=\ln\frac{1-p_{e}}{p_{e}}$ where $p_{e}$ is the (marginal) probability of Pauli error on that edge, edges $e\in \varepsilon$ supporting erasures have weight $w_e=0$, and the total
weight of a correction $\bar{l}$ is $w_{\bar{Z}}(\bar{l})=\sum_{e\in\bar{l}}w_{e}$. In reality, the error $\epsilon$ as part of $E$ is unknown and therefore which correction is correct is unknown; hence we only have access to the \textit{unsigned} logical gap, which we refer to simply as the \textit{logical gap}  $|\Delta_{\bar{Z}}(E)|$ (below we will drop the dependence on $E$ for brevity). 

In general, any decoder can be used to compute a logical gap and biased noise can be accommodated by modifying the weights appropriately. If one chooses a minimum-weight perfect-matching (MWPM) decoder,
then the decoder will always choose the minimum weight correction. If $\Delta_{\bar{Z}}<0$, the decoder will fail and a logical
error will be introduced. If $\Delta_{\bar{Z}}>0$,
the decoder will succeed in correcting the error and if $\Delta_{\bar{Z}}=0$, the decoder will succeed/fail half of the time. Therefore, the EER for the block becomes
\begin{equation}
    p_\mathrm{enc}=\sum_{i\in(\bar{Z},\bar{X})}\int_{-\infty}^{0}P(\Delta_{i})d\Delta_{i},
\end{equation}
where $P(\Delta_{i})$ is the distribution of logical gaps of logical membranes $i$ for a fixed block size and error rate. In more complex \textit{logical blocks}~\cite{bombin2021logical} (i.e., surface code protocols/channels), there will be many logical membranes and so one can
compute a vector of logical gaps as the soft-information of interest.

\begin{align}
    Q_{\mathrm{G},i}(v_{E})\coloneqq|\Delta_{i}|,i\in C,
\end{align}
where recall, $C$ is the set of distinct logical membranes. We can create a combined score for the block to be thresholded by the policy as

\begin{align}
    S_{\mathrm{G}}(Q_{\mathrm{G}})&\coloneqq\sum_{i\in C}a_{i}e^{-Q_{\mathrm{G},i}} \nonumber\\
    P_{\mathrm{G}}(Q_{\mathrm{G}};s_{\mathrm{G}}^{*})&\coloneqq\Theta(s_{\mathrm{G}}^{*}-S_{\mathrm{G}}(Q_{\mathrm{G}}))
\end{align}
where $a_{i}$ represent tunable linear weights to add the scores
of all logical membranes.

\subsubsection{Nested Logical Gap}  \label{subsubsec:NestedLogicalGap}

The nested logical gap rule $R_{\mathrm{N}}=(Q_{\mathrm{N}},P_{\mathrm{N}})$
is a derivative of the logical gap rule which combines information of the annular syndrome as the soft-information of interest.

\begin{align}
    Q_{\mathrm{N}}(v_{E};\alpha)\coloneqq(Q_{\mathrm{G}}(v_{E}),Q_{\mathrm{S}}(v_{E};\alpha)).
\end{align}
The policy is given by conditional thresholding expressed as

\begin{align}
       &P_{\mathrm{N}}(Q_{\mathrm{N}};s_{\mathrm{G}}^{*},s_{\mathrm{S}}^{*})= \begin{cases}
       1 ~&\text{ if }  S_{\mathrm{G}}(Q_{\mathrm{G}})=s_{\mathrm{G}}^{*} \text{ and } \\ & S_{\mathrm{S}}(Q_{\mathrm{S}})\leq s_{\mathrm{S}}^{*}, \text{ or } \\
        & S_{\mathrm{G}}(Q_{\mathrm{G}})<s_{\mathrm{G}}^{*}\\
       0 ~&\text{otherwise}
       \end{cases}
\end{align}

If one were to imagine a scenario of choosing $M$ out of $N$ configurations,
then this policy amounts to sorting all $N$ configurations first
by the logical gap and then by annular syndromes, choosing the best
$M$ configurations in order. The intuition is that the preparation
block has constant distance of $2$ to flip logical sectors and so
up to normalization, the gap, for a single graph (primal or dual),
is bounded to $|\Delta|\leq2$, thus leading to a large degeneracy
as we will see in Sec.~\ref{subsec:RulePerformance}. The idea is to use the annular syndrome rule to break this degeneracy.

\subsubsection{Radial Logical Gap}  \label{subsubsec:RadialLogicalGap}

The radial logical gap rule $R_{\mathrm{RG}}=(Q_{\mathrm{RG}},P_{\mathrm{RG}})$
is a derivative of the logical gap rule that caters specifically to
the structure of the preparation block. The radial logical gap rule
computes the logical gap but with a radial power-law (similar to the
annular syndrome rule with same cutoff) reweighting of the edge weights such that $\tilde{w}_{i}\coloneqq\frac{w_{i}}{\min(r,\lceil\frac{3L_\mathrm{d}}{4}\rceil)^{\alpha}}$. This yields $\tilde{\Delta}_{i}\coloneqq\tilde{w}_{i}(\bar{l}_{\mathrm{wrong}})-\tilde{w}_{i}(\bar{l}_{\mathrm{correct}})$
and
\begin{align}
    Q_{\mathrm{RG},i}(v_{E})\coloneqq|\tilde{\Delta}_{i}|,i\in C.
\end{align}
We can create a combined score for the block to be thresholded by
the policy as
\begin{align}
    S_{\mathrm{RG}}(Q_{\mathrm{RG}})&\coloneqq\sum_{i\in C}a_{i}e^{-Q_{\mathrm{RG},i}} \nonumber\\
    P_{\mathrm{RG}}(Q_{\mathrm{RG}};s_{\mathrm{RG}}^{*})&\coloneqq\Theta(s_{\mathrm{RG}}^{*}-S_{\mathrm{RG}}(Q_{\mathrm{RG}})).
\end{align}
The intuition here is to break the aforementioned degeneracy of the logical gap in the preparation block by biasing the decoder to compute corrections \emph{away} from the preparation point (and into the bulk) so that some entropic contributions are, in a heuristic manner, included.

\section{Results and Discussion}  \label{sec:Results}

\subsection{Simulation Details}  \label{subsec:SimDetails}

We consider the magic state preparation block and error model described in
Sec.~\ref{sec:MagicStatePrep}. Specifically, we consider \textit{i.i.d.} bitflip and erasure errors with strength $p_{\mathrm{error}}$ and $p_{\mathrm{erasure}}$ on the $XX$ and $ZZ$ fusion outcomes, as well as single qubit $X$ and $Z$ measurement outcomes. We Monte-Carlo sample $n_\mathrm{trials}=10^5$
trials of preparation block configurations, apply each rule to all
samples, and selectively keep the best $\kappa$ fraction of them.
We assess the encoding error rate $p_{\mathrm{enc}}$ (recall Eq.~(\ref{eq:preperror})) of each
rule as a function of $\kappa$. On a real quantum computer, any desired
$\kappa$ can be achieved, on average, by running the policy in real-time
with appropriate choice of score cutoffs. The Minimum-weight Perfect-matching
(MWPM) decoder~~\cite{dennis2002topological,kolmogorov2009blossom} is used for both decoding and computing the logical
gap. Surviving distance computations can be performed using Djikstra's shortest path algorithms modified to keep track of multiplicities. Unless otherwise
stated, all linear weights $\{a_{i}\}$ for all rules are set to unity
in the spirit of being fully agnostic between primal and dual graphs.
For logical gap and surviving distance computations, each boundary is attached to an additional \emph{pseudosyndrome} vertex, with pairs of pseudosyndromes associated to like boundaries (e.g., primal-primal) lit up to change/flip the sector for logical correction on that respective graph.

We assume on the output port that all surface code stabilizers are measured noiselessly, allowing for a logical readout in each basis, i.e., while the output qubits themselves are subject to noise, we assume no measurement noise on the stabilizer measurements (see Ref.~\cite{bombin2021logical} for more details).

\subsection{Rule Performance}  \label{subsec:RulePerformance}

\begin{figure*}
	\centering

    \includegraphics[width=1.0\linewidth]{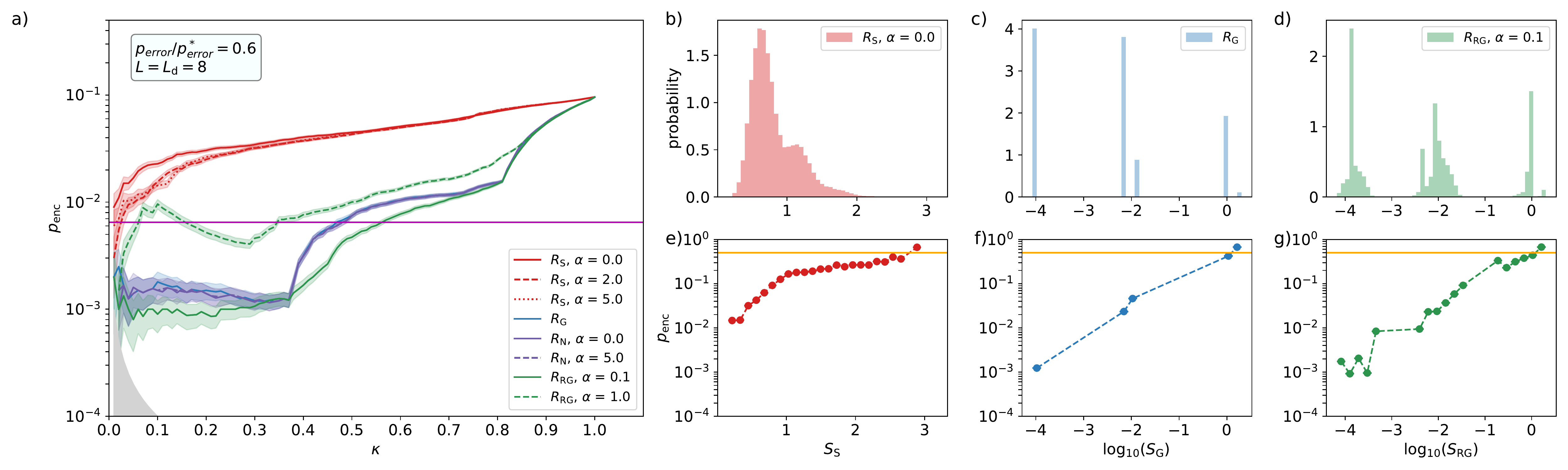}
	\caption{ (a) Encoding Error Rate (EER, $p_\mathrm{enc}$) of a $L=L_\mathrm{d}=8$ cubic magic state preparation block at $p_\mathrm{error}=0.6p_\mathrm{error}^*$, where $p_\mathrm{error}^*$ is the bulk threshold, for the annular syndrome (red), logical gap (blue), nested logical gap (purple), and radial logical gap (green) rules. The logical-gap-based rules give rapid suppression (step-like due to the discrete logical gap sectors seen in (c)) in the EER as compared the annular syndrome rule, with the radial gap rule at low power $\alpha=0.1$ performing the best. The magenta line indicates the ``breakeven" line where $p_\mathrm{enc}=p_\mathrm{init}=p_\mathrm{error}$, whereafter diminishing returns occur with decreasing $\kappa$ as per Eq. \ref{eq:preperror}. The gray region, where $p_\mathrm{enc}\leq\frac{1}{n_\mathrm{trials}\kappa}$, indicates the limits of sampling in the simulation. The shading around the colored lines denotes the standard error $(p_\mathrm{enc}(1-p_\mathrm{enc})/(n_\mathrm{trials}\kappa))^{1/2}$. (b,c,d) Distributions of scores for the annular syndrome ($S_\mathrm{S}$), logical gap ($S_\mathrm{G}$), and radial logical gap ($S_\mathrm{RG}$) rules, respectively. Annular syndrome scores have a continuous distribution whereas the logical gap sectors are discrete. The radial logical gap score at $\alpha=0.1$ weakly breaks the degeneracy of the logical gap scores. (e,f,g) The correlation of EER and score for the annular syndrome, logical gap, and radial logical gap rules, respectively. The orange line is at $p_\mathrm{enc}=0.5$ indicating the absence of correlation (uniform probability of either logical sector). The annular syndrome score has a poor correlation with EER while the logical gap and radial gap rules have a strong exponential correlation with EER thereby endowing predictive power to the logical gap. The radial logical gap rule, at $\alpha=0.1$, has a more continuous distribution/smoother correlation with the EER, thus improving upon the logical gap rule by smoothing out the ``step-like" features in (a).
	}
	\label{fig:rule_main}
\end{figure*}

We show the performance of all rules\footnote{The logical gap generalizes the surviving distance rule defined in App.~\ref{app:survivingdist} which is designed for erasure errors. The gap rule effectively subsumes the distance rule since it also incorporates syndromes resulting from Pauli errors, as seen in Figs.~\ref{fig:rule_main}, \ref{fig:rule_perror}, \ref{fig:rule_mixed_1:1}, and \ref{fig:rule_mixed_9:1}, and therefore is not shown.} in Fig.~\ref{fig:rule_main}a, as a function
of the keep fraction $\kappa$ for pure Pauli error
$p_{\mathrm{error}}=0.6p_{\mathrm{error}}^{*}$ and $p_{\mathrm{erasure}}=0$, where $p_{\mathrm{error}}^{*}=0.0108$
is the bulk threshold of the memory block. In other words, we show $p_{\mathrm{enc}}^{R}(\kappa;p_{\mathrm{error}}=0.6p_{\mathrm{error}}^{*},p_{\mathrm{erasure}}=0)$
for $R\in\{R_{\mathrm{S}},R_{\mathrm{G}},R_{\mathrm{N}},R_{\mathrm{RG}}\}$.
At $\kappa=1$, there is no post-selection and hence all rules have
the same EER, i.e., the same $p_\mathrm{enc}$. As $\kappa$ decreases and fewer blocks are accepted,
all rules suppress the EER, albeit at different rates. The overhead $O$ for postselection is equal to $1/\kappa$. If one assumes the same error rate on the initial magic state such that $p_{\mathrm{init}}=p_{\mathrm{error}}$,
then the intersection of the EER of each rule with the magenta line of Fig.~\ref{fig:rule_main} indicates the ``breakeven" keep value $\kappa^{*}(R)$ (or overhead $O^*=1/\kappa^*$) at which the EER is the
equal to the initial magic state error. As per Eq.~(\ref{eq:preperror}),
post-selection yields diminishing returns for $\kappa<\kappa^{*}(R)$
as the overall error rate $p_{\mathrm{prep}}$ becomes dominated by
$p_{\mathrm{init}}$ in this regime.

For $p_{\mathrm{error}}<p_{\mathrm{error}}^{*}$ as in Fig.~\ref{fig:rule_main}a, the EER suppression is super exponential (in $\kappa$) for the gap rule and its variants, with an
overhead of $O^*\lesssim2$, below which, there are diminishing returns, as the initial magic state error $p_{\mathrm{init}}$ will become the dominant source of error. When the differential overhead cost is low, i.e., $\frac{d\ln p_{\mathrm{enc}}^{R}}{d\kappa}\gg0$ in the regime around $\kappa^{*}$, it might desirable to use (relatively small) extra overhead to suppress the EER further below the initial magic state error. In contrast, in the same regime, the annular syndrome
rule $R_{\mathrm{S}}$ has poor suppression of the EER since the syndrome fraction is only loosely correlated with the EER. As shown in Appendix \ref{app:paulirules}, as $L$ increases, the annular syndrome rule performs increasingly poorly
since statistical fluctuations of obtaining finite size samples with
few syndromes are exponentially suppressed (cf. Fig.~\ref{fig:rule_main}b). However, for larger $L$, the gap rules all still perform well as the gap is effectively utilizing a
decoder rather than being reliant on statistical fluctuations at finite
size. For $p_{\mathrm{error}}\approx p_{\mathrm{error}}^{*}$ (at the bulk threshold value), as shown in Appendix \ref{app:paulirules}, interestingly the same qualitative behavior holds, but quantitatively the EER suppression rate is reduced as expected when the system is inherently more noisy.

To understand the performance differences between the various gap
rules, it is instructive to analyze the distributions of scores and the correlations of the scores with their respective EER. From Fig.~\ref{fig:rule_main}c, the distribution of
gap rule scores $S_{\mathrm{G}}(Q_{\mathrm{G}})$ is highly discrete/degenerate
due to the fact the gap for each logical membrane is bounded by the
constant fault distance of the preparation block $(|\Delta_{\mathrm{primal}}|,|\Delta_{\mathrm{dual}}|\in\{0,1,2\}$) and hence the combined score can only take on five distinct values,  $S_{\mathrm{G}}(Q_{\mathrm{G}})\in\{0,1,2,3,4\},\forall E$, up to normalization. As $p_{\mathrm{error}}$ increases, the distribution of gaps for concentrates around $\Delta = 0$. This results in an overall shift of the gap score distribution towards higher values indicating the configurations are typically more error-prone (due to decreased distinguishability between logical sectors -- see App.~\ref{app:paulidists} for more details). From Fig.~\ref{fig:rule_main}f, a decreasing gap score has an exponentially smaller EER and this strong correlation of the score and EER gives rise to the predictive power of the gap rule. Furthermore, the fact that the gap rule does not have an exponentially vanishing number of configurations at low scores is what also makes it practical with low overhead. 

In hopes of improving upon the gap rule by breaking the degeneracy of the gap sectors to yield a more fine-grained score, one can nest the annular syndrome score inside each of the
discrete gap sectors and assess the performance of this nested rule $R_{\mathrm{N}}$. This does provide minor improvements over certain ranges of $\kappa$ as compared to the gap rule (Fig.~\ref{fig:rule_main}a and Fig.~\ref{fig:rule_perror}), but is not particularly remarkable due to the poor correlation of syndromes with EER (Fig.~\ref{fig:rule_main}e) persisting inside each gap sector. In contrast, modifying the gap by adding an inverse radial
weighting in $R_{\mathrm{RG}}$ breaks the degeneracy of the gap scores (Fig.~\ref{fig:rule_main}d) by favoring corrections---that determine the gap---away from the initial magic state (where the fault distance is constant). For low power $\alpha=0.1$, the radial gap score $S_{\mathrm{RG}}(Q_{\mathrm{RG}})$ weakly breaks the degeneracy of the gap score while still preserving the gap sectors. This heuristically incorporates more entropic effects in the bulk, providing a smoother graded correlation of the radial gap score and the EER (Fig.~\ref{fig:rule_main}g), thus leading to improved predictive power of $R_{\mathrm{RG}}$ over $R_{\mathrm{G}}$. We find that $R_{\mathrm{RG}}$ is at least as good if not better than $R_{\mathrm{G}}$ for all $\kappa$. At higher $\alpha$, e.g., $\alpha=1.0$ as in Fig.~\ref{fig:rule_main}a, the radial gap rule performance degrades due to the now strong power-law which mixes gap sectors thus obtaining a poor, non-monotonic, correlation with the EER (see Appendix \ref{app:paulicorr}) and hence losing the original predictive power of the gap.

\begin{figure*}[t]
	\centering
    \includegraphics[width=0.4\linewidth]{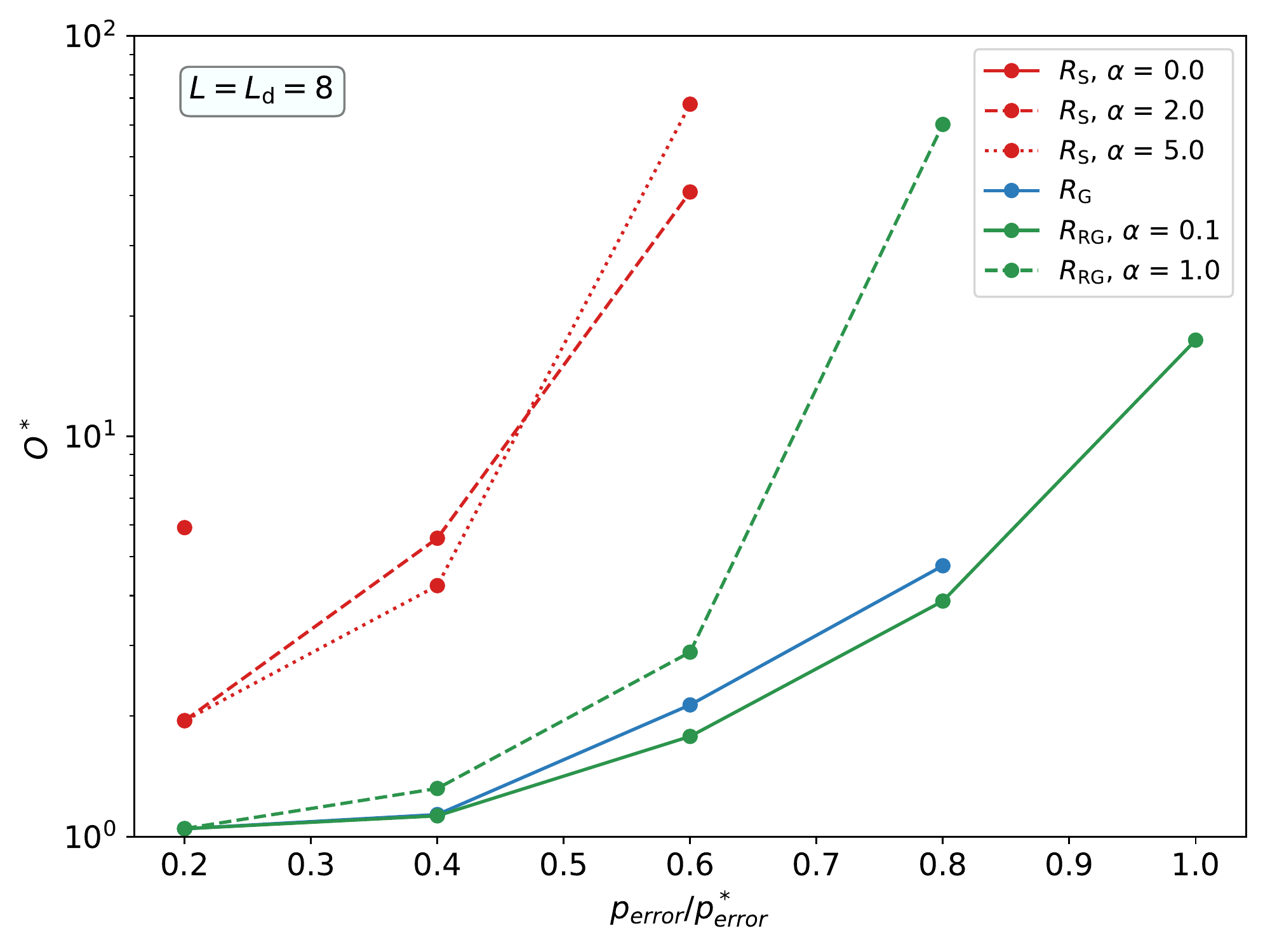} \quad
    \includegraphics[width=0.5\linewidth]{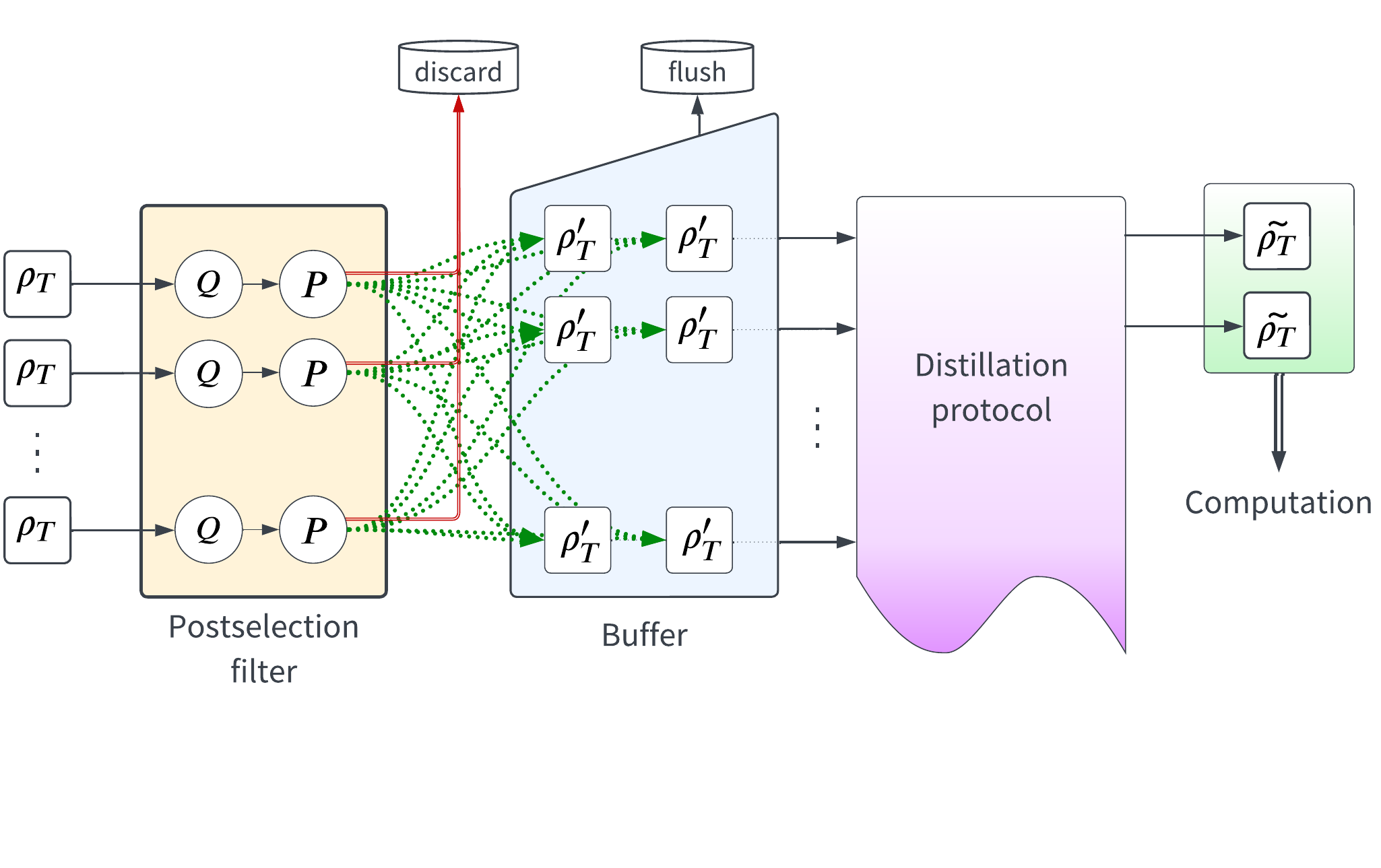}
	\caption{(left) Breakeven Overhead $O^*=\frac{1}{\kappa^{*}}$ as a function of the
fraction of the bulk threshold $p_{\mathrm{error}}/p_{\mathrm{error}}^{*}$ at $L=L_\mathrm{d}=8$ for different post-selection rules.
The annular syndrome rule has relatively poor performance with large overhead needed (as a function of error rate \emph{and} $L$) to achieve the breakeven point, suggesting that simple syndrome-counting based rules are inefficient. In stark contrast, the radial gap rule (green), at $\alpha=0.1$, outperforms the logical gap (blue), and annular syndrome (red) rules
by having the lowest overhead to achieve the breakeven condition $p_\mathrm{enc}=p_\mathrm{init}=p_\mathrm{error}$ over the entire error range. In particular, the radial the gap rule has a benign scaling for error rates below the bulk threshold with, for example, only a modest $1.78$ overhead at $0.6p_{\mathrm{error}}^{*}$ to reach the breakeven point. The nested gap rule has similar performance to the gap rule as seen and is therefore not shown for clarity. (right) Magic states are prepared, post-selected, and then stored in a buffer for distillation. In the figure, the initial encoded magic states $\rho_{T}$ are prepared and post-selected upon using the soft information function $Q$ and policy $P$. Magic states are either discarded (if rejected by the policy) or accepted, and if accepted they are sent to the buffer. Accepted magic states are denoted $\rho_{T}'$ and have error rate $p_{\mathrm{prep}}$ and are utilized in the distillation protocol. The output states of the distillation protocol are denoted $\tilde{\rho}_T$ and are used for fault-tolerant quantum computation. }
	\label{fig:Overhead}
\end{figure*}

To compare the performance of the rules over a range of Pauli error rates, in Fig.~\ref{fig:Overhead} (left) we show the breakeven overhead as a function of the
fraction of the bulk threshold $p_{\mathrm{error}}/p_{\mathrm{error}}^{*}$.
Over this entire range, the radial gap rule at low $\alpha$ has the lowest overhead required to reach the breakeven point, everywhere performing better than the gap and nested gap. In contrast, the annular
syndrome rule performs poorly. At an error rate of $p_{\mathrm{error}}=0.6p_{\mathrm{error}}^{*}$, the radial gap rule has relative overhead of only $1.78$, which is $\sim 23$ \emph{times} lower than the best annular syndrome rule and $1.17$ \emph{times} lower than the gap rule. As the error rate increases, the annular syndrome breakeven overhead increases exponentially due to its reliance on statistical fluctuations (e.g., configurations with zero syndromes are desirable but are exponentially rare) and quickly surpasses tractable simulation, hence the absence of breakeven points at higher errors in Fig.~\ref{fig:Overhead} (left). This is similarly seen, even for the gap rule with an absence of a breakeven point in the current simulation at $p_{\mathrm{error}}=p_{\mathrm{error}}^{*}$. On the flip side, however, even at the bulk threshold error rate, the radial gap rule can still achieve the breakeven condition at a reasonable overhead of $\sim17$. It is important to note that for a given rule, there may not always be a breakeven point, even with infinite simulation capacity, since above the optimal decoding threshold (in the thermodynamic limit) the visible information cannot be used to reliably distinguish logical sectors. These qualitative results hold true even in the presence of nonzero erasure as shown in Appendix \ref{app:erasures}.

\subsection{Architectural Design}  \label{subsec:ArchDesign}

In reality, the input magic state preparation blocks must be selected in real-time from a finite set. Further, several magic states are required for each round of distillation, and so one must determine how many parallel preparation sites---called \textit{preparation factories}---are required such that there is a sufficient rate of initial magic states reaching the first level of distillation. We propose and analyze a simple buffer-based
architecture to obtain a more accurate estimate of the cost and performance of the post-selection rules proposed in the previous sections. This buffer architecture is particularly well-suited to photonic FBQC architectures, but is applicable to matter-based CBQC architectures, provided the routing costs are  accounted for.

Consider $n_{\mathrm{fac}}$ preparation factories, each of which synchronously generate a magic state block on a clock with time interval $t_{\mathrm{fac}}$. Consider also a collective memory buffer that can store a number of magic state blocks for a time $t_{\mathrm{flush}}=n_{\mathrm{cycles}}t_{\mathrm{fac}}$, measured in the number of factory clock cycles $n_{\mathrm{cycles}}$, before the entire buffer, i.e., \emph{all} of its magic states, is erased. For a distillation protocol that takes in $m_{\mathrm{in}}$
blocks and outputs $m_{\mathrm{out}}$ blocks, if the buffer is not filled with $m_{\mathrm{in}}$ magic state blocks by $t_{\mathrm{flush}}$, distillation cannot proceed leading to wasted resources when the buffer is flushed. We assume that the temporal overhead for the classical computation needed for post-selection is negligible in-between the factories and the collective buffer, that there is all-to-all connectivity between factories and memory slots in the buffer as shown in Fig.~\ref{fig:Overhead} (right), that routing magic states is free (in space and time), as in Ref.~\cite{bombin2021interleaving}, and that all magic state factories are uncorrelated in terms of quality of initial magic states.

For a given post-selection rule $R$, each keep ratio $\kappa$ corresponds to a cutoff score(s) we more explicitly denote $s^{*}(\kappa;R)$ (for score $S_\mathrm{R}(Q)$), determined by numerical simulation apriori. At each factory clock cycle, $n_{\mathrm{fac}}$ magic state blocks are produced. A classical computational filter then applies the rule's policy on each block, only accepting a block if $S_\mathrm{R}(Q)\leq s^{*}(\kappa;R)$. The accepted blocks are moved into the buffer, and since the probability of accepting a single magic state block is by construction $\kappa$, there are on average $\kappa n_{\mathrm{fac}}$ blocks stored in the buffer after one clock cycle. Since the magic states produced in each clock cycle are uncorrelated with those produced in previous cycles, the collection of accepted
magic state blocks after $n_{\mathrm{cycles}}$ follows a binomial distribution with mean $\mu=n_{\mathrm{cycles}}n_{\mathrm{fac}}\kappa$ and variance $\sigma^{2}=n_{\mathrm{cycles}}n_{\mathrm{fac}}\kappa(1-\kappa)$. To ensure a filled buffer of size $m_\mathrm{in}$ up to failure probability $p_\mathrm{flush}$ for an acceptance probability $\kappa$, we solve $p_{\mathrm{flush}}=F(m_{\mathrm{in}}-1;n_{\mathrm{cycles}}n_{\mathrm{fac}},\kappa)$
for $n_{\mathrm{cycles}}n_{\mathrm{fac}}$, where $F(x; n, p)$ is the cumulative distribution function (cdf) for the binomial distribution of $n$ trials and success probability $p$. Note that $p_\mathrm{flush}$ rapidly decays in the regime of interest where $m_\mathrm{in}-1 < \mu$, and that the solution allows for a simple space-time tradeoff between $n_{\mathrm{cycles}}$ and $n_{\mathrm{fac}}$ (since the product remains fixed) which is useful for working around any physical resource constraints that might be present. Furthermore, $\frac{\sigma}{\mu}\sim\frac{1}{\sqrt{n_{\mathrm{cycles}}n_{\mathrm{fac}}}}$
means that the relative fluctuations of buffer filling vanish with larger magic state requirements, as would be the case for multiple rounds of distillation. Practical implementation of this collective buffer scheme with a common
flush time, or a variation that allows for individual flush times
rather than a collective flush time, both require a detailed specification
of a physical architecture and its description of errors that will
inform, for example, constraints on total overhead and constraints
on space-time geometries for routing magic state blocks.

A distillation protocol produces magic states with output error rate $f(p_\mathrm{prep};c,k)=cp_{\mathrm{prep}}^k$ (to first order), for some constants $c, k$, assuming the input magic state error rate $p_{\mathrm{prep}}$ is sufficiently small. For example, the well-known 15-to-1 distillation protocol~\cite{bravyi2005universal, litinski2019magic} outputs 1 magic state of quality arbitrarily close to $35p_{\mathrm{prep}}^3$ using 15 input magic states of quality $p_{\mathrm{prep}}$. This assumes the code distances for the input surface codes are large such that errors in the Clifford operations are negligible. For an algorithm of interest, with $n_T$ $T$-gates and $n_Q$ number of qubits, one must distill magic states of a error rate $p_{\mathrm{alg}} = O(\frac{1}{n_T n_Q})$ to run the entire algorithm with constant error rate. To achieve this, one must choose the distillation protocol such that $f(p_\mathrm{prep};c,k) < p_{\mathrm{alg}}$. We would like to choose the distillation protocol that achieves this output error rate with the fewest resource states possible. One should jointly optimize the distillation protocol (across the landscape of possible distillation protocols~\cite{haah2017magic,haah2017magicintermediate,haah2018codes}) and postselection protocol (i.e., the post-selection rule and how many preparation factories are required) to minimize overall resources. As we have seen, the radial gap rule achieves the lowest error rates to prepare magic states for a given postselection overhead (i.e., fixed $O$).

\section{Conclusions and Future Work}  \label{sec:Conclusions}

As fault-tolerant demonstrations on current quantum technologies are becoming more prevalent~\cite{doi:10.1126/science.abi8378,egan2021fault,ryan2021realization,postler2021demonstration, acharya2022suppressing}, it is essential to develop more accurate modeling and resource estimation tools to determine the requirements for large scale quantum computations. We have established a framework for fault-tolerant post-selection and applied it to the magic state problem, a dominant source of overhead for fault-tolerant quantum computations.
Our numerical results demonstrate that the post-selection rules we propose rapidly suppress the encoding error rate of initial magic states in surface code blocks---under an error model of \emph{i.i.d.} Pauli errors and erasure errors, and over a wide range of error rates---to the level of the initial magic state error, all for low constant multiplicative overhead of $\sim1.5-5$ times the cost of a single magic state preparation block. In particular, the logical gap---a post-selection rule inspired by the statistical mechanics-to-quantum error correction correspondence---and its variants serve as powerful soft-information metric at the topological level. We observe up to a $\sim 25$ times reduction of overhead compared to commonly used syndrome-based post-selection strategies, for practical operational regimes. The proposed gap-based post-selection protocols are general, and can be applied to a variety of different fault-tolerant primitives, including for example, the modified preparation protocols of Ref.~\cite{gavriel2022transversal}. This soft-information can inspire and serve as a foundation for post-selection rules and/or multiplexing strategies for other logical blocks as part of the error requirements in a larger quantum architectural stack.

Further reduction of space-time volume might be possible by reducing the depth of the preparation block - for example, preparation factories producing $(L,L_\mathrm{d})=(4,2)$ for post-selection might be sufficient for efficiently choosing quality blocks that can be routed into buffers, i.e., fused into large depth identity (memory) blocks. This short-depth situation is difficult to model as considered here (where post-selection and decoding both take place on the full information of the block) since the final time-like boundary layer of perfect measurements in simulation is a large fraction of the block. For other logical blocks, fault-tolerant protocols, and error models (such as correlated or biased), incorporating more information about the noise model and geometry into the logical gap computation may lead to further improvements.

\section*{Acknowledgments}
We thank Naomi Nickerson, Ye-Hua Liu, and Chris Dawson, for detailed feedback during the course of the project, and Sara Bartolucci, Patrick Birchall, Hugo Cable, Axel Dahlberg, Andrew Doherty, Dan Dries, Megan Durney, Terry Farrelly, Mercedes Gimeno-Segovia, Eric Johnston, Konrad Kieling, Isaac Kim, Daniel Litinksi, Sam Morley-Short, Andrea Olivo, Sam Pallister, Fernando Pastawski, William Pol, Terry Rudolph, Jake Smith, Chris Sparrow, Mark Steudtner, Jordan Sullivan, David Tuckett, Andrzej Perez Veitia, and all our colleagues at PsiQuantum for useful discussions.

\textbf{Contributions.} KS, SR, and MP developed the general fault-tolerant post-selection framework. HB is responsible for the conception and initial investigation of the logical gap. KS and SR are responsible for initial investigations into FTPS of logical blocks, the further development and analysis of the logical gap, conception and development of its variants and other rules, and writing of the manuscript. KS is responsible for building the majority of the simulation.


%

\section{Appendix: Surviving distance rule}\label{app:survivingdist}

The surviving distance rule $R_{\mathrm{D}}=(Q_{\mathrm{D}},P_{\mathrm{D}})$
relies solely on erasure information and computes an analog of the code distance of the block that remains after removing the erased clusters. Namely, we compute the length of the shortest path on the syndrome graph between two distinct boundaries, where erased edges have zero cost. In other words, the surviving distance is the minimal number of Pauli errors that can result in a logical error, given the observed erasure. Furthermore, we may augment
this rule by incorporating the multiplicity $m(d_{i})$ of the shortest
path into the scoring to compute an effective distance for each pair
of boundaries as the soft-information of interest.

\begin{align}
    Q_{\mathrm{D},i}(v_{E};c)\coloneqq d_{i}-c\ln m(d_{i}),i\in\mathrm{boundary\;pairs}
\end{align}
where $c$ is a tunable parameter that governs the weighting of the
multiplicity. Each separated
pair of boundaries (e.g., the two logical operators on a single surface
code memory block) contributes an effective distance and we can create
a combined score for the block to be thresholded by the policy as

\begin{align}
S_{\mathrm{D}}(Q_{\mathrm{D}}) & \coloneqq\sum_{i\in\mathrm{boundary\;pairs}}a_{i}e^{-Q_{\mathrm{D},i}} \nonumber\\
P_{\mathrm{D}}(Q_{\mathrm{D}};s_{\mathrm{D}}^{*})&\coloneqq\Theta(s_{\mathrm{D}}^{*}-S_{\mathrm{D}}(Q_{\mathrm{D}}))
\end{align}
where $a_{i}$ are tunable linear weights to add the scores of different
pairs of boundaries. The preparation block only has 2 pairs of boundaries
terminating the $\bar{X},\bar{Z}$ logical correlator and $i=\mathrm{primal,\;dual}$.
The intuition is that spanning paths between boundaries mimic logical
errors chains and so having smaller effective distances due to erasure
make the configuration less desirable.

\section{Appendix: Detailed error model analysis}\label{app:fullsimresults}

We show extensive numerical results for the performance of the rules discussed in the main text over a range of Pauli error and erasure error as discussed in Sec.~\ref{subsec:ErrorModel}. In the case of pure Pauli errors, we also include the score distributions and the correlations of the EER and the scores for several different choices of rule parameters.

\subsection{Pauli Errors} \label{app:pauli}

\subsubsection{Rule Performance} \label{app:paulirules}

\begin{figure*}[t]
	\centering

    \includegraphics[width=1.0\linewidth]{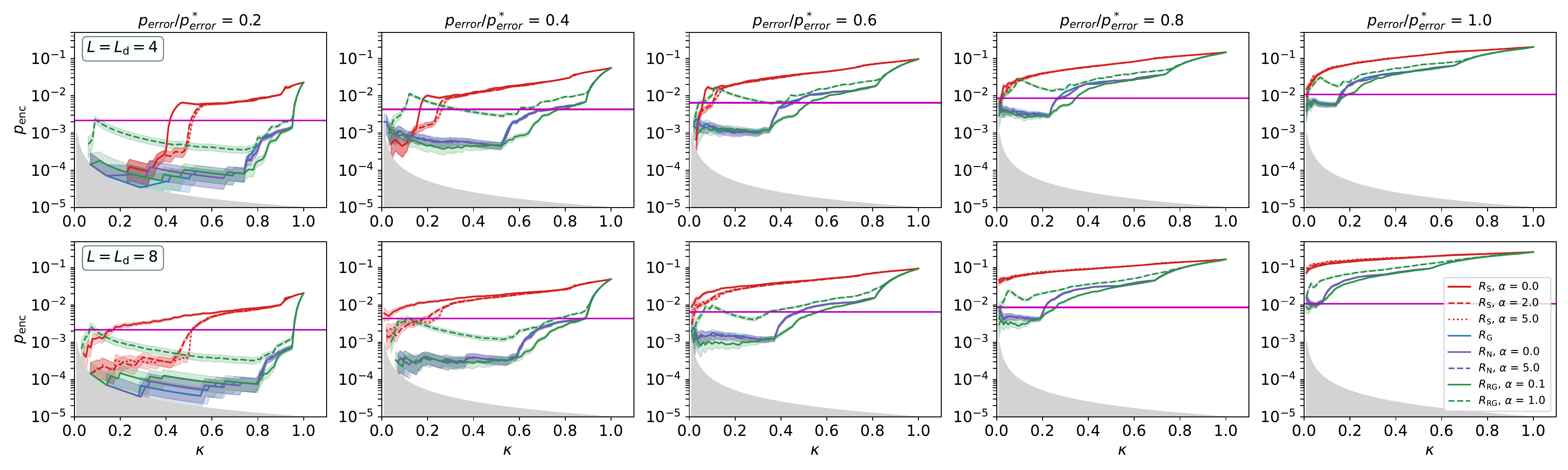}
	\caption{Encoding Error Rate of a $L=L_\mathrm{d}=4,8$ (first and second row, respectively) cubic magic state preparation block over a range Pauli error rate $p_\mathrm{error}/p_\mathrm{error}^*\in[0,1]$ for the annular syndrome, logical gap, nested logical gap, and radial logical gap rules. The shading around the colored lines denotes the standard error $(p_\mathrm{enc}(1-p_\mathrm{enc})/(n_\mathrm{trials}\kappa))^{1/2}$.
	}
	\label{fig:rule_perror}
\end{figure*}

The encoding error rate as a function of keep fraction is shown in Fig.~\ref{fig:rule_perror} for select values of $p_\mathrm{error}/p_\mathrm{error}^*\in[0,1]$. The first row shows results at $L=L_\mathrm{d}=4$ and the second row shows results at $L=L_\mathrm{d}=8$. Note that the all rules perform qualitatively similarly to that shown in the main text, with the expected degradation of performance as $p_\mathrm{error}$ approaches the bulk threshold $p_\mathrm{error}^*$. It is interesting to note, that even at threshold, at the smaller size where finite size effects are strong and beneficial, one can still hit the breakeven point with a relative overhead of approximately $6$ to $7$.

\subsubsection{Score distributions} \label{app:paulidists}

\begin{figure*}
	\centering

    \includegraphics[width=1.0\linewidth]{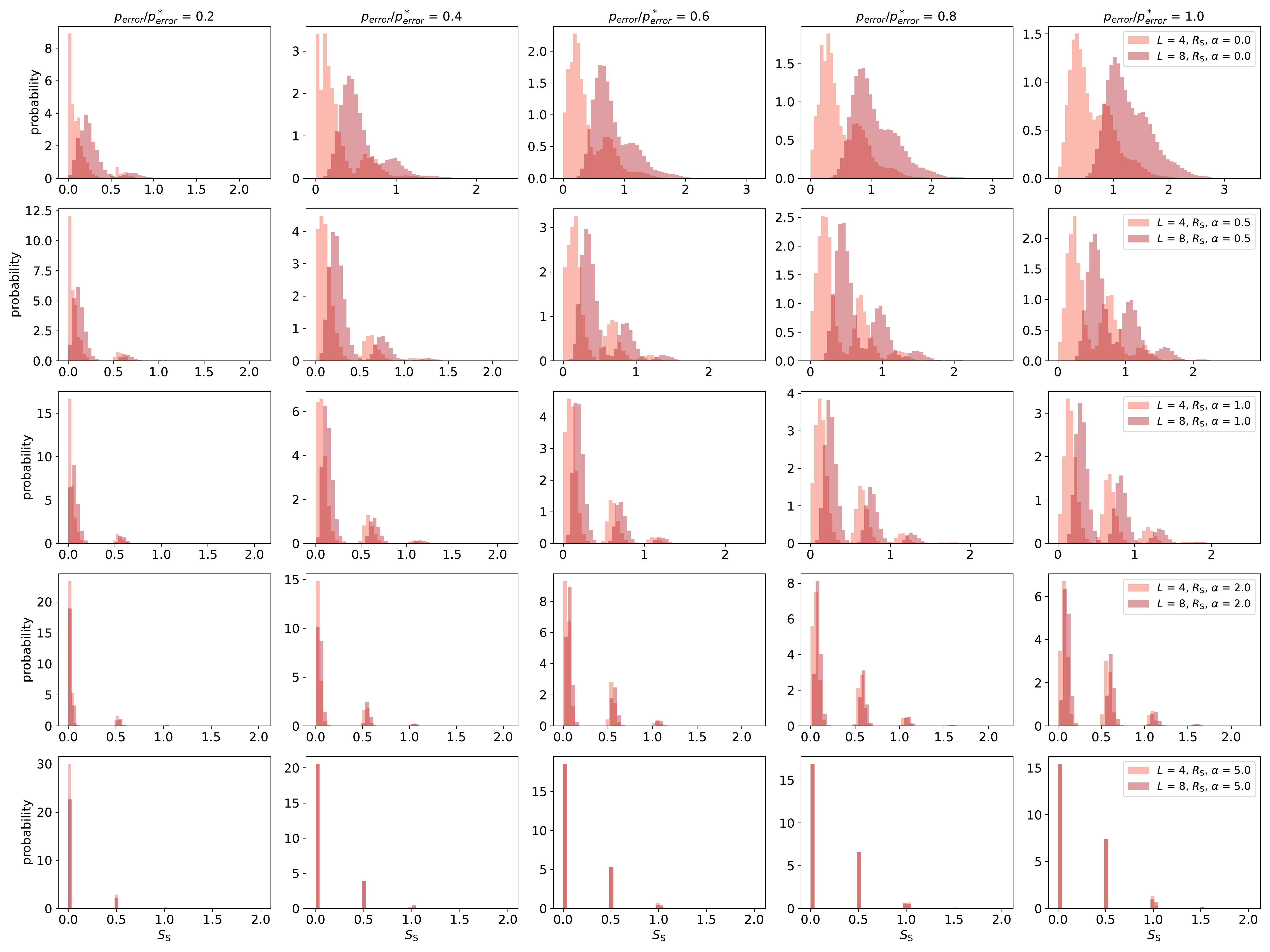}
	\caption{Score distribution for the annular syndrome rule. Each column has a fixed Pauli error rate in the range $p_\mathrm{error}/p_\mathrm{error}^*\in[0,1]$. Each row has a fixed choice of the rule parameter $\alpha$, the power-law decay exponent of the radial weighting.
	}
	\label{fig:softsyn_hist_perror}
\end{figure*}

\begin{figure*}
	\centering

    \includegraphics[width=1.0\linewidth]{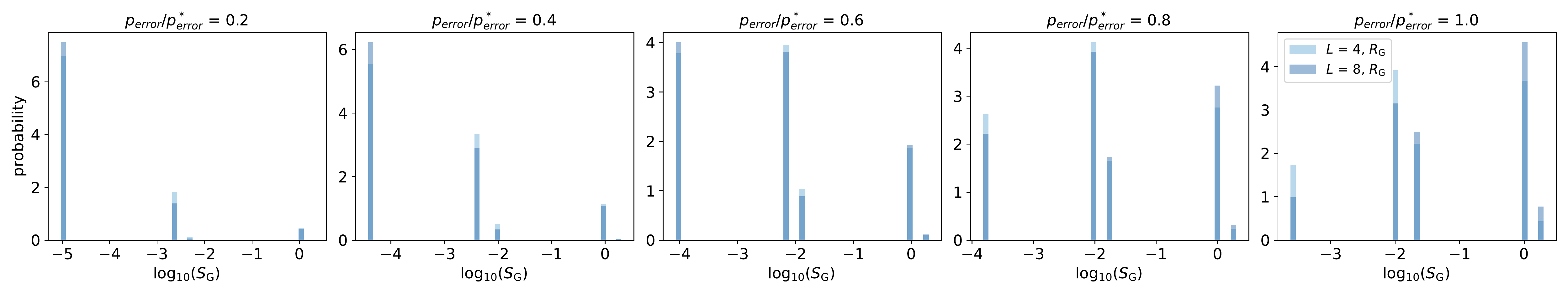}
	\caption{Score distribution for the gap rule over Pauli error rates in the range $p_\mathrm{error}/p_\mathrm{error}^*\in[0,1]$.
	}
	\label{fig:gap_hist_perror}
\end{figure*}

\begin{figure*}
	\centering

    \includegraphics[width=1.0\linewidth]{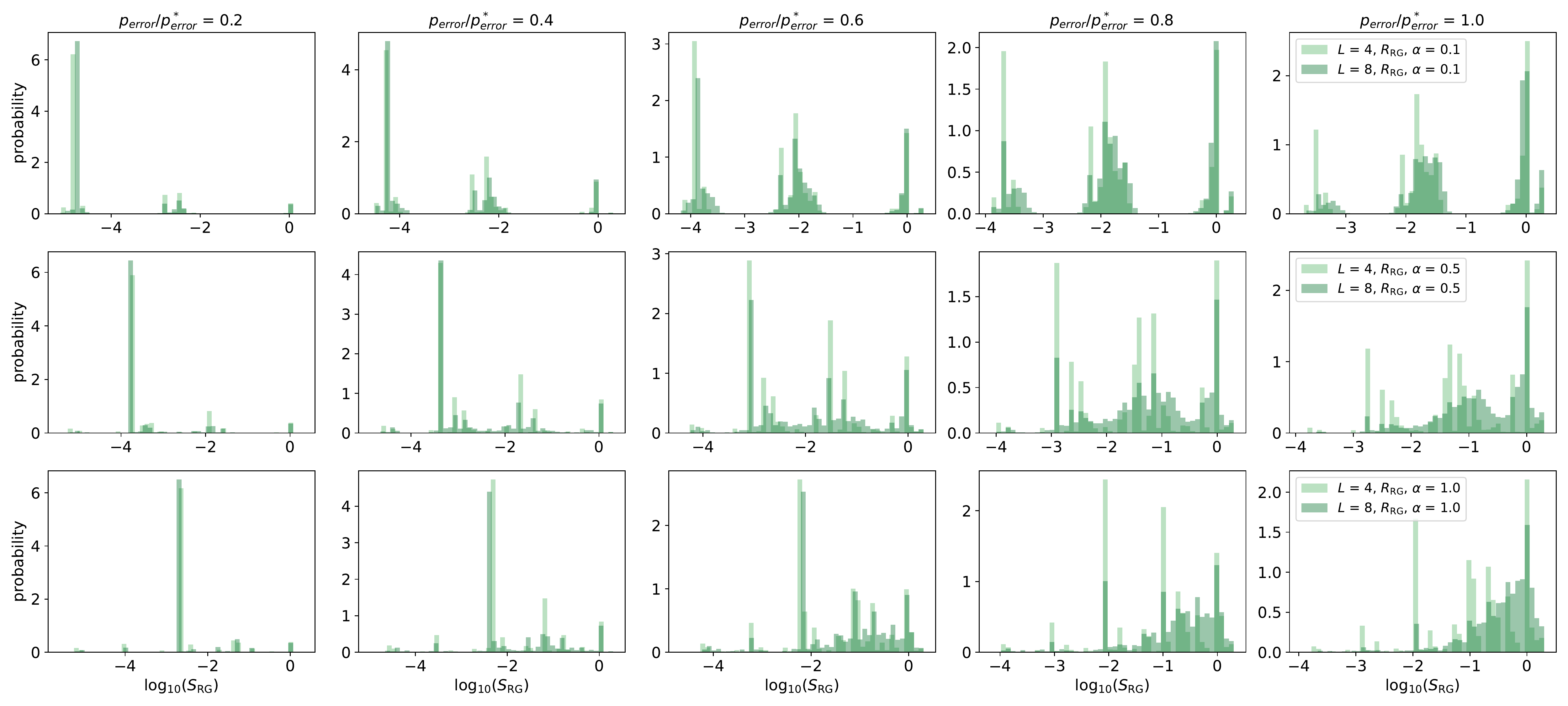}
	\caption{Score distribution for the radial gap rule. Each column has a fixed Pauli error rate in the range $p_\mathrm{error}/p_\mathrm{error}^*\in[0,1]$. Each row has a fixed choice of the rule parameter $\alpha$, the power-law decay exponent of the radial weighting.
	}
	\label{fig:radial_gap_hist_perror}
\end{figure*}

In Fig.~\ref{fig:softsyn_hist_perror}, Fig.~\ref{fig:gap_hist_perror}, and Fig.~\ref{fig:radial_gap_hist_perror}, we show the distributions of scores for the annular syndrome, logical gap, and radial logical gap rules, over a range of rule parameters (each row) and for $p_\mathrm{error}/p_\mathrm{error}^*\in[0,1]$ (each column), respectively. Increasing the power-law exponent $\alpha$ in the annular syndrome rule squeezes the distribution of scores, which apparently leads to performance improvement as seen in Fig.~\ref{fig:rule_perror}. As discussed in the main text, the gap rule has a discrete distribution over the full range of errors. In the radial gap rule, a small power-law exponent $\alpha$ mildly breaks this degeneracy and spreads the gap sectors while a large $\alpha$ mixes and reorganizes the gap sectors entirely, with the former yielding superior rule performance as seen in Fig.~\ref{fig:rule_perror}.

\subsubsection{Correlation of EER and Scores} \label{app:paulicorr}

\begin{figure*}
	\centering

    \includegraphics[width=1.0\linewidth]{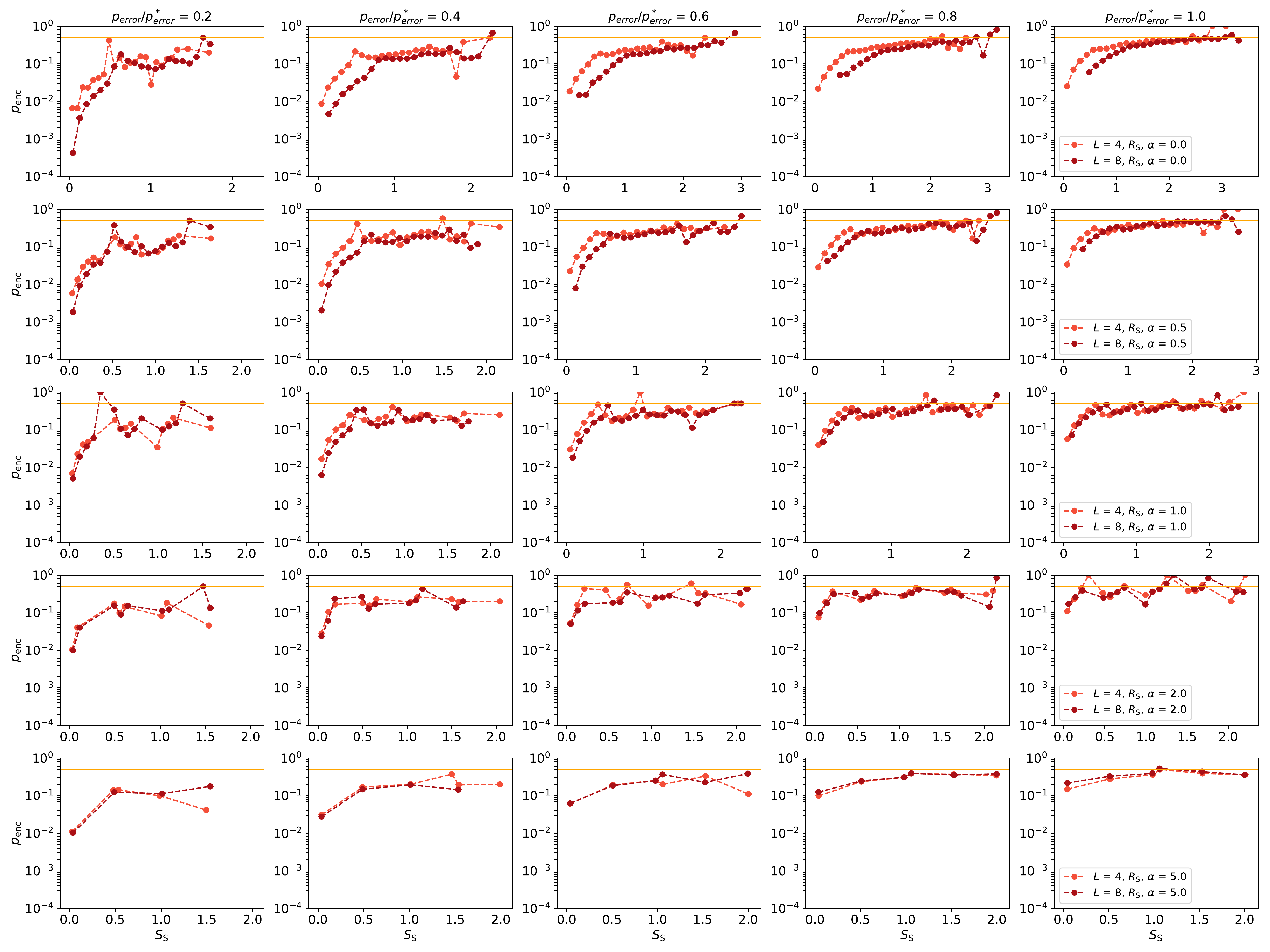}
	\caption{Correlation of the EER and the score for the annular syndrome rule. Each column has a fixed Pauli error rate in the range $p_\mathrm{error}/p_\mathrm{error}^*\in[0,1]$. Each row has a fixed choice of the rule parameter $\alpha$, the power-law decay exponent of the radial weighting.
	}
	\label{fig:softsyn_corr_perror}
\end{figure*}

\begin{figure*}
	\centering

    \includegraphics[width=1.0\linewidth]{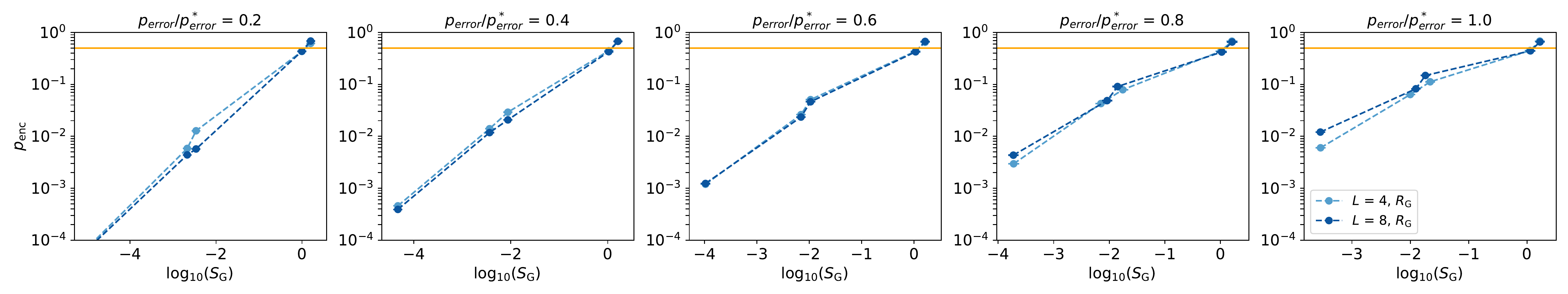}
	\caption{Correlation of the EER and the score for the gap rule over range of Pauli error rate in the range $p_\mathrm{error}/p_\mathrm{error}^*\in[0,1]$.
	}
	\label{fig:gap_corr_perror}
\end{figure*}

\begin{figure*}
	\centering

    \includegraphics[width=1.0\linewidth]{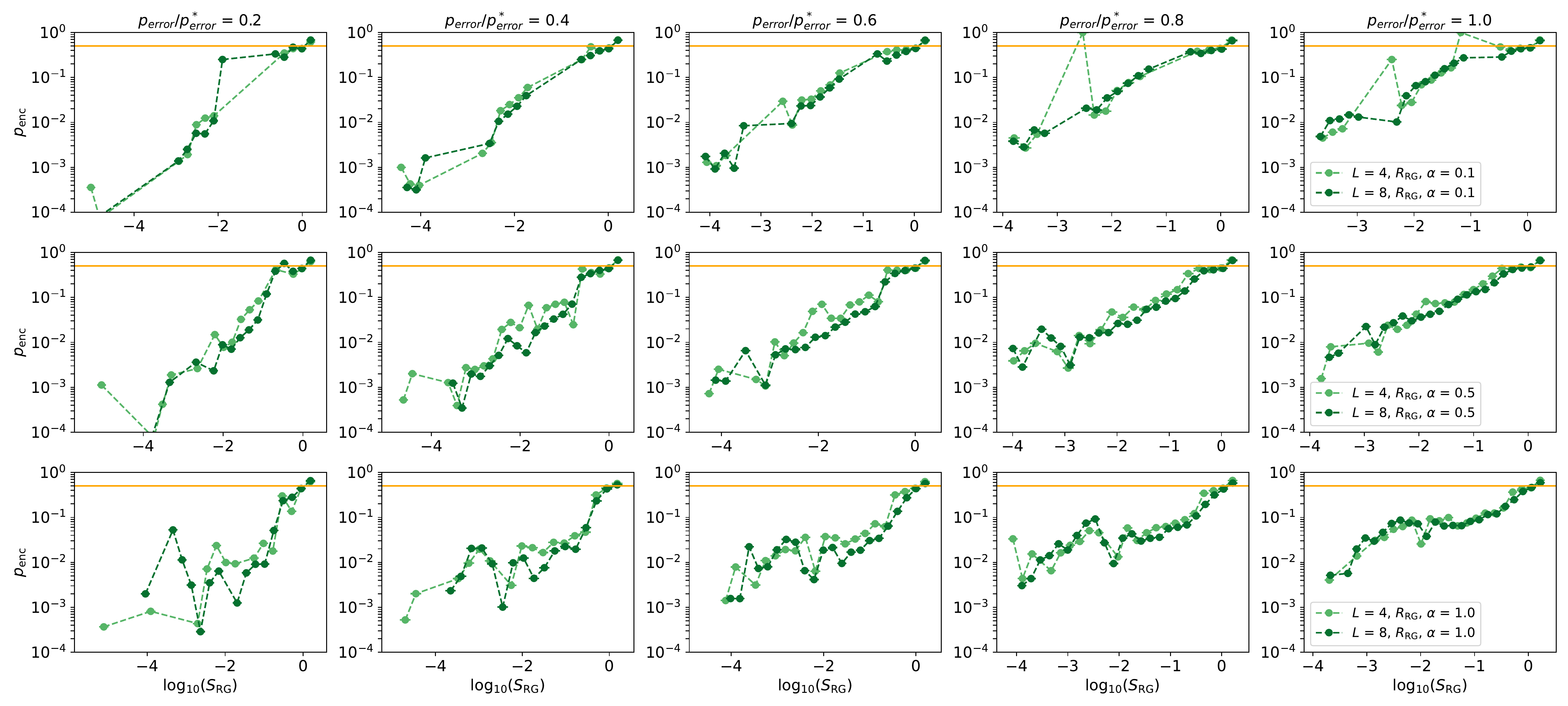}
	\caption{Correlation of the EER and the score for the radial gap rule. Each column has a fixed Pauli error rate in the range $p_\mathrm{error}/p_\mathrm{error}^*\in[0,1]$. Each row has a fixed choice of the rule parameter $\alpha$, the power-law decay exponent of the radial weighting.
	}
	\label{fig:radial_gap_corr_perror}
\end{figure*}

In Fig.~\ref{fig:softsyn_corr_perror}, Fig.~\ref{fig:gap_corr_perror}, and Fig.~\ref{fig:radial_gap_corr_perror}, we show the correlations of scores and the EER for the annular syndrome, logical gap, and radial logical gap rules, over a range of rule parameters (each row) and for $p_\mathrm{error}/p_\mathrm{error}^*\in[0,1]$ (each column), respectively. As before, the correlation of annular syndrome score and EER is weak, for all $\alpha$, although for low values it is stronger and hence beneficial. Again, the gap and radial gap have a strong correlation of their respective scores and the EER, leading to the significant performance gains as discussed in the main text, with the low $\alpha$ of the radial gap leading to a more monotonic and consistent correlation thus yielding the best performance.

\subsection{Mixed Erasure and Pauli Errors} \label{app:erasures}

\begin{figure*}
	\centering

    \includegraphics[width=1.0\linewidth]{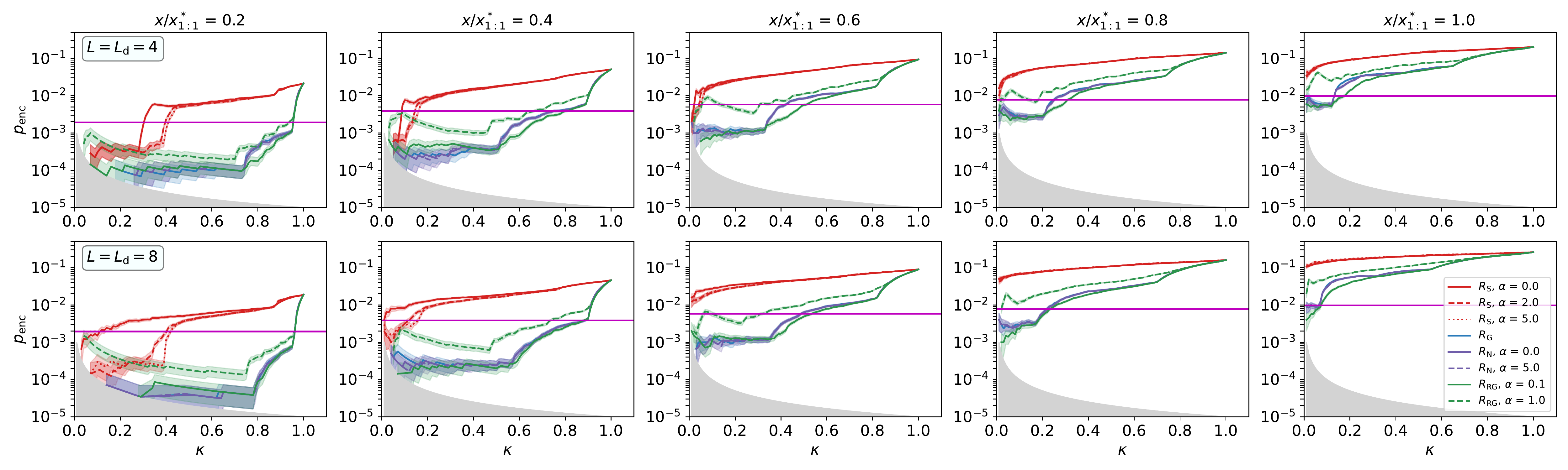}
	\caption{Encoding Error Rate of a $L=L_\mathrm{d}=8$ cubic magic state preparation block over a mixed erasure/Pauli error model with $(p_\mathrm{erasure}, p_\mathrm{error})=(x,x)$ where $x^*_{1:1}=9.71\times10^{-3}$ for the annular syndrome, logical gap, nested logical gap, and radial logical gap rules. The shading around the colored lines denotes the standard error $(p_\mathrm{enc}(1-p_\mathrm{enc})/(n_\mathrm{trials}\kappa))^{1/2}$.
	}
	\label{fig:rule_mixed_1:1}
\end{figure*}

\begin{figure*}
	\centering

    \includegraphics[width=1.0\linewidth]{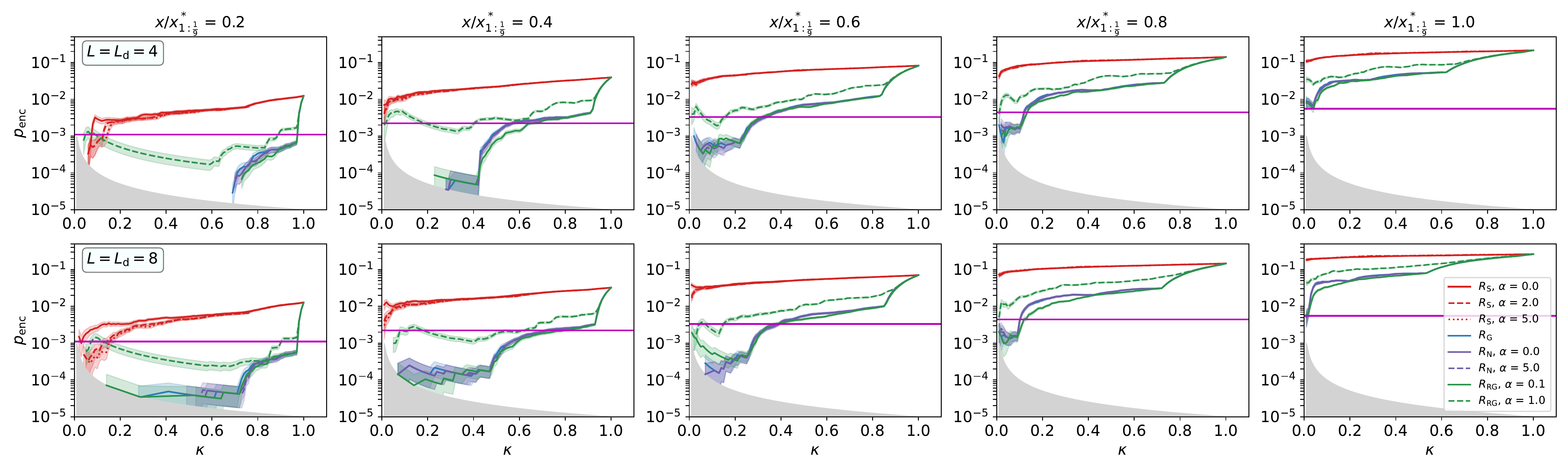}
	\caption{Encoding Error Rate of a $L=L_\mathrm{d}=8$ cubic magic state preparation block over a mixed erasure/Pauli error model with $(p_\mathrm{erasure}, p_\mathrm{error})=(x,\frac{x}{9})$ where $x^*_{1:\frac{1}{9}}=4.99\times10^{-2}$ for the annular syndrome, logical gap, nested logical gap, and radial logical gap rules. The shading around the colored lines denotes the standard error $(p_\mathrm{enc}(1-p_\mathrm{enc})/(n_\mathrm{trials}\kappa))^{1/2}$.
	}
	\label{fig:rule_mixed_9:1}
\end{figure*}

We consider both erasure and Pauli errors. For this error model, fusion outcomes are erased with probability $p_\mathrm{erasure}$, and non-erased outcomes are further subject to a bitflip outcome with rate $p_\mathrm{error}$. We simulate two representative cases with ($p_\mathrm{erasure}, p_\mathrm{error})=(x,x)$ and $(p_\mathrm{erasure},p_\mathrm{error})=(x,\frac{x}{9})$, where $x/x^*\in[0,1]$ and $x^*$ is the bulk threshold along the error ray parametrized by $x$. This threshold is determined empirically in both cases with MWPM decoding as $x^*_{1:1}=9.71\times10^{-3}$ and $x^*_{1:\frac{1}{9}}=4.99\times10^{-2}$, respectively. The results for the $1{:}1$ case are shown in Fig.~\ref{fig:rule_mixed_1:1} and the results for the $1{:}\frac{1}{9}$ case are shown in Fig.~\ref{fig:rule_mixed_9:1}, the latter being a more physically relevant scenario for FBQC using photonics, where loss (which leads to fusion outcome erasure) is a dominant source of error. In both cases, the qualitative behavior is similar to that of the pure Pauli error case, demonstrating that the gap-based post-selection rules yield significant improvement in the presence of erasures as well.

\end{document}